\documentclass[12pt,pre,aps,preprint,showpacs]{revtex4}

\usepackage{graphics}
\usepackage{graphicx}
\usepackage{amsfonts}
\usepackage{amsmath}
\usepackage{float}
\usepackage{epsfig}

\begin{document}
\title{Effect of Dimensionality on the Percolation Threshold of Overlapping Nonspherical Hyperparticles}

\author{S. Torquato}

\email{torquato@princeton.edu}

\affiliation{\emph{Department of Chemistry, Department of Physics,
Princeton Institute for the Science and Technology of
Materials, and Program in Applied and Computational Mathematics}, \emph{Princeton University},
Princeton NJ 08544}

\author{Y. Jiao}

\email{yjiao@princeton.edu}

\affiliation{\emph{Princeton Institute for the Science and Technology of
Materials}, \emph{Princeton University},
Princeton NJ 08544}

\begin{abstract}
In order to study the effect of dimensionality on the percolation threshold $\eta_c$ of 
overlapping nonspherical convex hyperparticles
with random orientations in $d$-dimensional Euclidean
space $\mathbb{R}^d$, we formulate a scaling relation for $\eta_c$ that is based on a rigorous 
lower bound [S. Torquato, J. Chem. Phys. {\bf 136}, 054106 (2012)] and a 
conjecture that hyperspheres provide the highest threshold
among all convex hyperparticle shapes for any $d$.
This scaling relation also exploits the recently discovered principle that  low-dimensional continuum percolation
behavior encodes high-dimensional information.
We derive an explicit formula for the  exclusion volume $v_{\mbox{\scriptsize ex}}$ of a hyperparticle
of arbitrary shape in terms of its $d$-dimensional volume $v$, surface area $s$ and {\it 
radius of mean curvature} ${\bar R}$ (or, equivalently, {\it mean width}). These basic geometrical
properties are computed for a wide variety  of nonspherical hyperparticle shapes with random orientations across all dimensions,
including, among other shapes, various polygons for $d=2$, Platonic solids, spherocylinders, parallepipeds
and zero-volume plates for $d=3$ and their appropriate generalizations for $d \ge 4$.
Using this information, we compute the lower bound and 
scaling relation for $\eta_c$ for this comprehensive set of 
continuum percolation models across dimensions. We demonstrate that the scaling relation provides accurate
{\it upper-bound} estimates of the threshold $\eta_c$ across dimensions and becomes increasingly accurate
as the space dimension increases.


\end{abstract}

\pacs{64.60.ah, 05.70.Jk, 71.30.+h, 82.70.Dd}

\maketitle
\section{Introduction}

Clustering and percolation behavior of many-particle systems is of relevance
in a host of physical and chemical phenomena, including  gelation and polymerization,
structure of liquids and glasses, hopping in semiconductors, metal-insulator transition in condensed matter systems,
nucleation, condensation of gases, chemical association, conduction in dispersions,
aggregation of colloids, and flow in porous media \cite{Za83,Br90,St92,Sa94,To02a,Bross}.
Considerable attention has been devoted to percolation on lattices, 
whose percolation thresholds, in some two-dimensional cases, are amenable
to exact analysis  \cite{Sa94, To02a}. Unlike lattice percolation, 
it has not been possible to obtain 
exact results for the percolation threshold 
of any two-dimensional continuum (off-lattice) percolation model.
Recently, analytical estimates of
the threshold, including  rigorous bounds, have been obtained for the prototypical continuum
percolation models of overlapping hyperspheres
and oriented hypercubes in $d$-dimensional Euclidean
space $\mathbb{R}^d$ that apply in  any dimension $d$ \cite{To12a, paper2}.
In this paper, we provide an analytical framework to predict
percolation thresholds for a wide class of continuum percolation
models, namely, overlapping hyperparticles in $\mathbb{R}^d$ of any convex shape 
and in any dimension $d$.

An important consequence of the analysis in Ref \cite{To12a}, which we exploit 
in the present work, is that the large-$d$ percolation value for any
hyperparticle shape is an important contribution to the
corresponding low-dimensional percolation value. In other words,
low-dimensional percolation properties encode high-dimensional information. The
analysis was aided by a remarkable {\it duality} between the
equilibrium hard-hypersphere (hypercube) fluid system and the
continuum percolation model of overlapping hyperspheres
(hypercubes), namely,
\begin{equation}
P(r;\eta)=-h(r;-\eta),
\label{dual}
\end{equation}
where $P(r;\eta)$ is the pair connectedness function at some
radial distance $r$, the quantity
\begin{equation}
\eta= \rho v
\label{den}
\end{equation}
is a dimensionless or reduced density, $\rho$ is the number density and $v$ is the volume of a
hyperparticle \cite{sub}. Additionally, $h(r;\eta)$ is the total correlation
function for the corresponding equilibrium hard-particle models at
{\it packing fraction} $\eta$, fraction of space covered by the hard
particles.  The duality relation (\ref{dual}) relates a statistical property of a topological problem on the one hand
to that of a geometrical problem on the other \cite{duality}. Importantly, the duality relation (\ref{dual}) is exact
in the large-$d$ limit and for low densities for any $d$ up to through
the third virial level, and is a relatively accurate approximation in  low dimensions for 
densities up to the percolation threshold $\eta_c$ \cite{To12a}.

It was  also shown in Ref. \onlinecite{To12a} that lower-order Pad{\' e}
approximants of the mean cluster number $S$ of the form $[n,1]$
(where $n=0,1$ and 2) provide upper bounds on $S$ and hence yield
corresponding lower bounds on the threshold $\eta_c$ for
$d$-dimensional overlapping hyperspheres and overlapping oriented
convex hyperparticles with central symmetry \cite{central} (e.g., spheres, cubes,
regular octahedra, and regular icosahedra for $d=3$) \cite{To12a}.
The simplest of these lower bounds on $\eta_c$ ($[0,1]$ Pad{\' e}
approximant) is given by
\begin{equation}
\eta_c \ge \frac{1}{2^d},
\label{isot}
\end{equation}
where it is to be noted that $2^d$ is the ratio of the exclusion
volume $v_{\mbox{\scriptsize ex}}$ associated oriented convex
hyperparticle \cite{exclude} (which includes the hypersphere) to the volume $v$
of the hyperparticle. This trivially translates into a lower bound
on the mean number of overlaps per sphere at the threshold ${\cal
N}_c$:
\begin{equation}
{\cal N}_c \equiv 2^d \eta_c \ge 1.
\end{equation}
The  lower bound estimates as well as Percus-Yevick-like
approximations were demonstrated \cite{To12a} to become accurate
in relatively low dimensions, improve in accuracy as $d$
increases, and become asymptotically exact in the large-$d$ limit,
i.e.,
\begin{equation}
\eta_c \rightarrow \frac{1}{2^d}, \qquad d \rightarrow \infty
\label{asy}
\end{equation}
and
\begin{equation}
{\cal N}_c \sim 1, \qquad d \rightarrow \infty.
\end{equation}

The aforementioned trends and asymptotic results were shown
to hold for overlapping convex $d$-dimensional particles (or {\it hyperparticles})
of {\it arbitrary shape and orientational distribution}  when appropriately
generalized \cite{To12a}. For example, for identical overlapping
hyperparticles of general anisotropic shape of volume $v$ with
specified probability distribution of orientations in $d$
dimensions, the simplest lower bound on $\eta_c$ and on ${\cal
N}_c$ generalize as follows:
\begin{equation}
\eta_c \ge \frac{v}{v_{\mbox{\scriptsize ex}}},
\label{aniso}
\end{equation}
\begin{equation}
{\cal N}_c \equiv \eta_c \frac{v}{v_{\mbox{\scriptsize ex}}} \ge 1,
\label{N-aniso}
\end{equation}
where
\begin{equation}
\label{v_ave}
v_{\mbox{\scriptsize ex}}=\int_{\mathbb{R}^d} f({\bf r},{\boldsymbol \omega}) p({\boldsymbol \omega}) d{\bf r}d{\boldsymbol \omega},
\end{equation}
is the generalized {\it exclusion volume} associated with a hyperparticle, 
$ f({\bf r},{\boldsymbol \omega})$ is the exclusion-region
indicator function \cite{To12a}, ${\bf r}$ and ${\boldsymbol \omega}$ is the
centroid position and orientation of one particle, respectively,
with respect to a coordinate system at the centroid of the other
particle with some fixed orientation, and $p({\boldsymbol
\omega})$ is the orientational probability density function.
Moreover, in the high-dimensional limit for any convex hyperparticle
shape, the following exact asymptotic results \cite{N_c} immediately follow:
\begin{equation}
\label{aniso_bd} \eta_c \sim \frac{v}{v_{\mbox{\scriptsize ex}}},
\qquad d \rightarrow \infty
\end{equation}
and
\begin{equation}
\label{aniso_bd2} {\cal N}_c \sim 1, \qquad d \rightarrow \infty.
\end{equation}
It was demonstrated in Ref. \onlinecite{To12a} that the lower bound (\ref{aniso})
on $\eta_c$ improves in accuracy in any fixed dimension as the particle
shape becomes more anisotropic. This property will be exploited later
in Sec. \ref{approx} to obtain an upper bound on $\eta_c$.

We note that the bounds (\ref{aniso}) and (\ref{N-aniso}),
and the associated asymptotic limits (\ref{aniso_bd}) and
(\ref{aniso_bd2}) also apply to overlapping {\it zero-volume} ($d-1$)-dimensional
{\it hyperplates} with specified orientations in $\mathbb{R}^d$ when
the relevant parameters are appropriately generalized. While these 
``infinitesimally thin'' hyperplates in $\mathbb{R}^d$ have vanishing volumes,
their  exclusion volumes are nonzero at number density $\rho$.
However, an appropriately defined ``effective volume" $v_{\mbox{\scriptsize eff} }$ must be chosen 
in order to define a reduced density $\eta$ of the form (\ref{den}) for hyperplates.
The choice for this effective hyperplate volume that we use
henceforth is the volume of a $d$-dimensional hypersphere of {\it characteristic radius} 
$r$, i.e.,
\begin{equation}
v_{\mbox{\scriptsize eff} }= \frac{\pi^{d/2}}{\Gamma(d/2+1)}r^d, \label{plate_vol}
\end{equation}
where $r$ is related to some characteristic length scale of the hyperplate
and $\Gamma(x)$ is the Euler-Gamma function. Thus, bounds (\ref{aniso}) and (\ref{N-aniso})
still apply for hyperplates when $v$ in expression for the reduced density $\eta$ is replaced
by the effective volume (\ref{plate_vol}), i.e.,
\begin{equation}
\eta_c \ge \frac{v_{\mbox{\scriptsize eff}}}{v_{\mbox{\scriptsize e}}}.
\label{aniso-2}
\end{equation}

Application of the aforementioned continuum percolation results for nonspherical hyperparticles was 
only briefly explored in Ref. \onlinecite{To12a}.  The main objective
of the present paper is to carry out such an investigation
for a wide variety  of nonspherical convex hyperparticles with {\it random orientations}
across dimensions in order to ascertain the effect of dimensionality
on the corresponding thresholds. This will be done
by first ascertaining the exclusion volume of a wide variety  of nonspherical hyperparticle shapes,
and then applying the lower bound (\ref{aniso}) on $\eta_c$ and a scaling relation for $\eta_c$,
the latter of which we obtain in the present paper. 

All of the results described above reveal that the topological
problem of percolation in low dimensions becomes a purely geometrical problem
in high dimensions. This behavior is explicitly manifested by the duality relation (\ref{dual})
and the fact the threshold $\eta_c$ is determined solely by the exclusion
volume associated with a hyperparticle in the high-$d$ limit [cf. (\ref{aniso_bd})],
a purely geometrical characteristic.
Exploiting the latter result, the principle that low-dimensional results encode high-dimensional information \cite{To12a},
and a conjecture (postulated here) that hyperspheres provide the highest threshold
among all convex hyperparticle shapes for any $d$,
enables us to devise the aforementioned scaling relation for $\eta_c$ for randomly oriented
hyperparticles of general nonspherical shape, which is also shown to  bound $\eta_c$
from above.

The rest of the paper is organized as follows. In Sec. \ref{random},
we derive an explicit formula for the  exclusion volume $v_{\mbox{\scriptsize ex}}$
associated with a nonspherical convex hyperparticle of arbitrary shape
in terms of its basic geometrical properties.
In Secs. \ref{2-3} and \ref{d}, we compute these geometrical properties for a wide variety  of 
nonspherical hyperparticle shapes with random orientations in two and three
 dimensions, and for $d \ge 4$.
A conjecture that systems of overlapping hyperspheres  provide the highest threshold
among all convex hyperparticle shapes for any $d$ is postulated in Sec. \ref{CONJ}.
In Sec. \ref{approx}, we derive a scaling relation for $\eta_c$ and show that
it is also an upper bound on $\eta_c$.
In Sec. \ref{appl}, we compute the lower bound (\ref{aniso}) and 
scaling relation for $\eta_c$ for the aforementioned comprehensive set of systems
of overlapping hyperparticles and compare the results to numerical
data for $\eta_c$, including those that we obtain here for the Platonic solids.
Finally, in Sec. \ref{conc}, we close with concluding remarks.

\section{Exclusion Volume for Randomly Oriented Overlapping Nonspherical Hyperparticles}
\label{random}

The exclusion volume $v_{\mbox{\scriptsize ex}}$ associated with 
randomly oriented overlapping convex nonspherical hyperparticles
plays a central role in this paper. In what follows, we derive an explicit formula for
$v_{\mbox{\scriptsize ex}}$
in terms of the $d$-dimensional volume $v$, surface area $s$ and {\it mean
radius of mean curvature} ${\bar R}$ (or {\it mean width}) of a hyperparticle
of arbitrary shape.

\subsection{Basic Geometrical Characteristics of a Convex Body}
\label{width}

\begin{figure}
\begin{center}
\includegraphics[height=3.5cm,keepaspectratio]{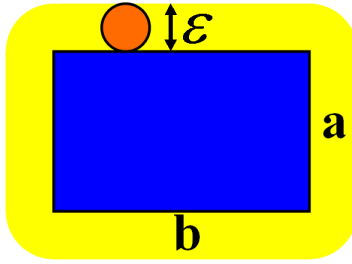}
\end{center}
\caption{(Color online) Two-dimensional illustration of the parallel body at a distance $\epsilon$ associated 
with a rectangle of side lengths $a$ and $b$, which is the union 
of the blue (or dark gray) and yellow (or light gray) regions.} 
\label{fig_para_body}
\end{figure}

Consider any convex body $K$ in $d$-dimensional Euclidean space
$\mathbb{R}^d$. Let $v$ and $s$ denote the $d$-dimensional volume
and surface area of $K$, respectively. For example, in
$\mathbb{R}^2$, $v$ and $s$ are the area and perimeter,
respectively, of the two-dimensional convex body or disk. In
Euclidean space $\mathbb{R}^d$, the parallel body associated with
the convex body $K$ at distance $\epsilon$ is equal to the sum of
$K$ and a Euclidean ball of diameter $\epsilon$. This operation
preserves the convexity and compactness properties associated with
$K$. The notion of a parallel body or set is well known in convex
geometry and has been applied in the physical sciences, even for
non-convex geometries; see, for example, the so-called {\it
dilation} processes that has been applied in the study of
heterogeneous materials \cite{To86i,To02a,Za11f}.
Figure \ref{fig_para_body} illustrates the parallel body associated with a rectangle in two 
dimensions.

The famous Steiner formula \cite{St95} expresses the volume of the parallel
body $v_\epsilon$ at distance $\epsilon$ as a polynomial in
$\epsilon$ and in terms of geometrical characteristics of the
convex body $K$, such as $v$ and $s$. For example, for the cases
$d=2$ and $d=3$, Steiner's formula yields respectively
\begin{equation}
v_\epsilon= v+ s \epsilon + \pi \epsilon^2
\label{Stein-2}
\end{equation}
and
\begin{equation}
v_\epsilon= v+ s \epsilon + 4\pi {\bar R} \epsilon^2 + \frac{4\pi \epsilon^3}{3},
\label{Stein-3}
\end{equation}
where
\begin{equation}
{\bar R}= \frac{1}{8\pi} \int \left (\frac{1}{R_1}+\frac{1}{R_2}\right)dS,
\label{m-w-0}
\end{equation}
is the {\it radius of mean curvature} of the convex body, $R_1$ and $R_2$ are the principle radii of curvature
and $dS$ denotes the integral over the entire surface of the convex body.

\begin{figure}
\begin{center}
\includegraphics[height=5.0cm,keepaspectratio]{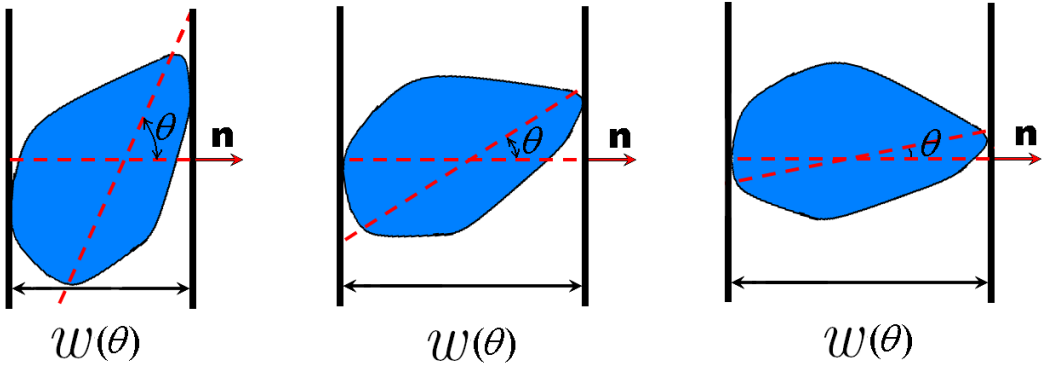}
\end{center}
\caption{(Color online) Illustration of the
width $w$ of a convex body in two dimensions. In $\mathbb{R}^2$, $w(\theta)$ is
function of the angle $\theta$ between the normal ${\bf n}$ of the
parallel planes (shown as red arrows) and the convex body. The
mean with $\bar w$ is obtained by averaging $w(\theta)$ over the
angle $\theta$, i.e., integrating $w(\theta)$ over the
angle $\theta$ and dividing by $2\pi$.}
\label{fig_mean_width}
\end{figure}

The Steiner formulas (\ref{Stein-2}) and (\ref{Stein-3})
can be generalized for a convex body $K$ in any dimension $d$:
\begin{equation}
v_\epsilon= \sum_{k=0} W_k \epsilon^k,
\label{Stein-d}
\end{equation}
where $W_k$ are trivially related to the {\it quermassintegrals} or {\it Minkowski functionals} \cite{St95}.
Note that $W_0=v$, $W_1=s$ and  $W_d= v_s(d;\epsilon)$, where
\begin{equation}
v_s(d;a)= \frac{ \pi^{d/2}a^d}{\Gamma(1+d/2)}
\label{sphere}
\end{equation}
is the volume of a $d$-dimensional sphere (hypersphere) of radius $a$. Of particular,
interest in this paper  is the {\it lineal} characteristic  in (\ref{Stein-d}), i.e., the $(d-1)$th coefficient:
\begin{equation}
\quad W_{d-1}= \Omega(d) {\bar R},
\label{m-w-1}
\end{equation}
where 
\begin{equation}
\Omega(d) = \frac{{d} \pi^{d/2}}{\Gamma(1+d/2)}
\label{solid}
\end{equation}
is the total solid angle contained in $d$-dimensional sphere and 
$\bar R$ is the radius mean of curvature in an space dimension $d$, which 
is trivially related to the {\it mean width} $\bar w$ of $K$ via
the following expression:
\begin{equation}
{\bar R}= \frac{\bar w}{2}.
\label{m-w}
\end{equation}
The mean width $\bar w$ is a {\it lineal} (one-dimensional)
measure of the ``size'' of $K$. Consider the convex body to be trapped
entirely between two impenetrable parallel ($d-1$)-dimensional
hyperplanes  that are orthogonal  to a  unit vector  ${\bf n}$ in
$\mathbb{R}^d$. The ``width" of a body $w({\bf n})$ in the
direction ${\bf n}$ is the distance between the closest pair of such
parallel hyperplanes, implying that the hyperplanes contact the
boundaries of the body. The mean width ${\bar w}$ is the average
of the width $w({\bf n})$ such that ${\bf n}$ is uniformly
distributed over the unit sphere $S^{d-1} \in \mathbb{R}^d$.
Figure \ref{fig_mean_width} illustrates the concept of the width of a convex body in
$\mathbb{R}^2$.

\subsection{Explicit Formula for the $d$-Dimensional Exclusion Volume}
\label{explicit}

General formulas for the exclusion volume associated with identical randomly
oriented nonspherical
convex particles in two and three dimensions have long been known \cite{Bo75,Kih53}. For example,
for $d=2$, it is known \cite{Bo75} that
\begin{equation}
v_{\mbox{\scriptsize ex}} = 2v+ \frac{s^2}{2\pi}
\label{2}
\end{equation}
where $s$ is the two-dimensional surface area or {\it perimeter} of the convex body.
Note that the first term of $2v$ is trivial, accounting for the fact that two bodies
are excluding one another and hence the exclusion volume must be at least
$2v$. At first glance, one might surmise that the appropriate three-dimensional
expression for the exclusion volume involves a second term that is proportional to $s^{3/2}/(4\pi)$,
where $s$ is the surface area of the three-dimensional convex body. However,
instead the correct three-dimensional formula \cite{Kih53}
is
\begin{equation}
v_{\mbox{\scriptsize ex}} = 2v+ \frac{s M}{2\pi},
\label{3}
\end{equation}
where
\begin{equation}
M= {4\pi} {\bar R}. 
\label{mean}
\end{equation}
The second terms in the two- and three-dimensional formulas
(\ref{2}) and (\ref{3}), respectively, appear to be functionally distinct from
one another. However, we will see shortly that by expressing both
formulas in terms of ${\bar R}$ a unifying formula emerges
that not only is valid for $d=2$ and $d=3$, but for arbitrary
$d$.


To our knowledge, a formula for the exclusion volume associated with identical randomly
oriented $d$-dimensional convex
nonspherical bodies $K$ for any $d$ has heretofore not been given explicitly. We find that this exclusion volume is
given by
\begin{equation}
v_{\mbox{\scriptsize ex}} = 2v+ \frac{2(2^{d-1}-1)}{d} s {\bar R}.
\label{vex}
\end{equation}
We arrive at this general formula for any $d$ by first recognizing that for the special case
$d=2$, the radius of mean curvature is trivially related to the two-dimensional
surface area or perimeter, namely, $s=2\pi{\bar R}$, which enables us to rewrite
relation (\ref{2}) as follows:
\begin{equation}
v_{\mbox{\scriptsize ex}} = 2v+ s {\bar R}.
\end{equation}
Similarly, using Eq. (\ref{mean}), we can re-express relation (\ref{3}) for $d=3$
in terms of $\bar R$, namely,
\begin{equation}
v_{\mbox{\scriptsize ex}} = 2v+  2s {\bar R}.
\end{equation}
We see that these two- and three-dimensional formulas differ by a coefficient multiplying
the product $s {\bar R}$, independent of the shape of the convex body.
Assuming that this coefficient depends only on dimension,  
the general $d$-dimensional formula for $v_{\mbox{\scriptsize ex}}$ associated
with any convex body $K$  must have the form
\begin{equation}
v_{\mbox{\scriptsize ex}} = 2v+  C(d)s {\bar R},
\label{gen}
\end{equation}
where $C(d)$ is a $d$-dimensional coefficient that still must be determined. 
To determine $C(d)$ as a function of $d$, we exploit the assumption that (\ref{gen}) 
must apply to any convex body so that one can use the most convenient
shape, i.e., a $d$-dimensional sphere (hypersphere). Note that the dimensionless exclusion volume 
$v_{\mbox{\scriptsize ex}}/v$ of a hypersphere, among all convex hyperparticles of nonzero volume, 
takes on its {\it minimum} value of $2^d$ \cite{To12a}.
For a hypersphere of radius $a$,  the volume $v$ is given by (\ref{sphere}), ${\bar R}=a$, and  the surface area
is given by
\begin{equation}
s_s(d;a)= \frac{d\pi^{d/2}}{\Gamma(1+\frac{d}{2})} a^{d-1}.
\label{s_s}
\end{equation}
Using these last four geometrical-property relations in conjunction with formula (\ref{gen}) yields
\begin{equation}
C(d) = \frac{2^d-2}{d}.
\end{equation}
Substitution of this expression  into relation (\ref{gen}) gives the general relation (\ref{vex}) for the
exclusion volume.

We now show  that the general formula (\ref{vex}) is indeed exact for any $d$.
Using relation (\ref{vex}) and the fact that $v_{\mbox{\scriptsize ex}}/v$ is minimized for 
hyperspheres (with $v_{\mbox{\scriptsize ex}}/v=2^d$) among all convex hyperparticles of nonzero volume, 
we can obtain the following
inequality involving $s$, ${\bar R}$ and $v$;
\begin{equation}
s {\bar R} \ge d v,
\label{iso}
\end{equation}
where the equality holds for hyperspheres only. This bound is a type
of {\it isoperimetric inequality} and has been proved in the study
of convex geometry using completely different techniques \cite{Sc93}.
The fact that the general expression (\ref{vex}) for $v_{\mbox{\scriptsize ex}}$ 
independently leads to a rigorously proven isoperimetric inequality for any convex 
body in $\mathbb{R}^d$ clearly justifies our assumptions concerning the form  
(\ref{gen}) and indirectly proves that (\ref{vex})
is an exact formula for any convex body in $\mathbb{R}^d$.
Moreover, we have computed $v_{\mbox{\scriptsize ex}}/v$ for 
a hypercube in $\mathbb{R}^4$ using Monte-Carlo simulations and verified
that the numerical value agrees very well with the analytical result 
given by (\ref{vex}), as expected.  

Note that for $d=2$, inequality (\ref{iso}) yields the standard
isoperimetric inequality $s \ge 4\pi v$, where we have used
the fact that the surface or perimeter $s= 2\pi {\bar R}$.
The $d$-dimensional generalization
of this standard isoperimetric inequality is 
\begin{equation}
s \;\Omega(d)^d \ge d v^{\frac{d-1}{d}},
\end{equation}
where $\Omega(d)$ is given by (\ref{solid})  \cite{Sc93}. It is clear that this inequality is
not the same as (\ref{iso}).

We see that in the case that a ($d-1)$-dimensional convex {\it hyperplate} 
in $\mathbb{R}^d$
that  possesses zero volume, i.e., $v=0$,
the exclusion-volume formula (\ref{vex}) yields
\begin{equation}
v_{\mbox{\scriptsize ex}} =  \frac{2(2^{d-1}-1)}{d}s {\bar R}.
\label{zero}
\end{equation}
We shall consider examples of such convex bodies for $ d\ge 3$
in the subsequent section.

\section{Radii of Mean Curvature and Exclusion Volumes for Various Nonspherical Particles
in Two and Three Space Dimensions}
\label{2-3}

Here we present results for the aforementioned geometrical characteristics
of a variety of convex bodies that are randomly oriented in two and three dimensions possessing 
both nonzero and zero volumes.
In the former instance, it is convenient to consider the dimensionless exclusion
volume $v_{\mbox{\scriptsize ex}}/v$.
Two- and three-dimensional exclusion volumes are known for a variety of convex nonspherical shapes that are randomly oriented \cite{Kih53,Bo75}. It  necessarily follows that such
exclusion volumes are larger
than the corresponding values when all of the particles are aligned
and centrally symmetric, in which case
$v_{\mbox{\scriptsize ex}}/v_1=2^d$, independent of the particle shape \cite{To12a}.

\subsection{Two Dimensions}

Table \ref{vol-2} provides volumes, surface areas, radii of mean curvature, and exclusion volumes
for some two-dimensional convex particles. This includes the circle, certain regular polygons, 
rectangle, spherocylinder and ellipse. The exclusion volumes for these particles
are easily be obtained from relation (\ref{2}). Alternatively, the  radius of mean
curvature $\bar R$ can be obtained from the volume of the parallel body associated with
the convex body at distance $\epsilon$ that is of order $\epsilon$ and use
of relation (\ref{m-w-1}). The corresponding exclusion volume can then be found
employing relation (\ref{vex}) for $d=2$.

In the ``needle-like" limit for an ellipse ($b/a \rightarrow \infty$), the general quantities $\bar R$ and $v_{\mbox{\scriptsize ex}}/v$
given in Table \ref{vol-2} become
\begin{equation}
{\bar R}\sim \frac{2b}{\pi}, \qquad \frac{v_{\mbox{\scriptsize ex}}}{v} \sim \frac{8}{\pi^2}\frac{b}{a}, \qquad b/a \rightarrow \infty.
\label{ellipse}
\end{equation}
The radius of mean curvature for  a randomly oriented, zero-volume {\it lines} of length $a$
in $\mathbb{R}^2$  is easily obtained from the rectangular case in Table \ref{vol-2} with $a_1=0$ and $a_2=a$,  yielding
\begin{equation}
{\bar R}=\frac{a}{\pi} \qquad \mbox{(line)}.
\label{line-1}
\end{equation}
The corresponding exclusion volume is extracted from  Table \ref{vol-2}
 in the rectangular case by first
multiplying both sides of the relation for $v_{\mbox{\scriptsize ex}}/v$
 by  the volume $v=ab$, and then
setting $a_1=0$ and $a_2=a$, which gives
\begin{equation}
v_{\mbox{\scriptsize ex}}=\frac{2a}{\pi} \qquad \mbox{(line)}.
\label{line-2}
\end{equation}
Not surprisingly, ${\bar R}$ for a line of length $a$ is identical to that of 
a needle-like ellipse of major axis $2b=a$ [cf. (\ref{ellipse})].

\begin{table}[H]
\caption{Volumes, surface areas, radii of mean curvature, and exclusion volumes
for some convex particles $K$ in $\mathbb{R}^2$. The radius of mean curvature
and exclusion volume correspond to randomly oriented convex
particles having nonzero volumes.}
\begin{center}
\begin{tabular}{c|c|c|c|c }
$K$ & $v$ & $s$ & ${\bar R}$ & $v_{\mbox{\scriptsize ex}}/v$   \\
\hline
{\scriptsize Circular disk, radius $a$} ~~&  ~~{\scriptsize $\pi a^2$}~~&~~{\scriptsize $2\pi a$} ~~&~~{\scriptsize  $a$}~~ &~~{\scriptsize $4$} \\
{\scriptsize Octagon, side $a$} ~~&~~{\scriptsize $2(1+\sqrt{2})a^2$}~~&~~{\scriptsize $8a$} ~~&~~$\frac{4a}{\pi}$~~ &~~  ${\scriptstyle 2}+\frac{16}{\pi(1+\sqrt{2})}$ \\
{\scriptsize Hexagon, side $a$} ~~&~~$\frac{\sqrt{3}a^2}{4}~~$~~&~~{\scriptsize $6a$} ~~&~~$\frac{3a}{\pi}$~~ &~~  ${\scriptstyle 2}+\frac{4\sqrt{3}}{\pi}$  \\
{\scriptsize Pentagon, side $a$} ~~&~~$\frac{\sqrt{25+10\sqrt{5}}a^2}{4}\;$  ~~&~~{\scriptsize $3a$} ~~&~~$\frac{5a}{2\pi}$~~ &~~  ${\scriptstyle 2}+\frac{50}{\pi\sqrt{25+10\sqrt{5}}}$  \\
{\scriptsize Square, side $a$} ~~&  ~~{\scriptsize $a^2$} ~~&~~{\scriptsize $4a$} ~~&~~$\frac{2a}{\pi}$~~ &~~${\scriptstyle 2}+ \frac{8}{\pi}$    \\
{\scriptsize Equilateral triangle, side $a$} ~~&~~$\frac{\sqrt{3}a^2}{4}$  ~~&~~{\scriptsize $3a$} ~~&~~$\frac{3a}{2\pi}$~~ &~~  ${\scriptstyle 2}+\frac{6\sqrt{3}}{\pi}$  \\
{\scriptsize Rectangle, sides $a_1$, $a_2$} ~~&  ~~{\scriptsize $a_1 a_2$}  ~~&~~{\scriptsize $2(a_1+a_2)$} ~~&~~$\frac{a_1+a_2}{\pi}$~~ &~~${\scriptstyle 2}+ \frac{2(a_1+a_2)^2}{a_1a_2\pi }$ \\
{\scriptsize Spherocylinder, radius $a$}~~& & & &\\
{\scriptsize cylindrical height $h$ }           &~ {\scriptsize $\pi a^2 +2ah$}~&~{\scriptsize $2(\pi a +h) $}~&~ $\frac{h}{\pi}+a$~& ${\scriptstyle 2}+\frac{2(\pi a+h)^2}{\pi a(\pi a+2h)}$\\
{\scriptsize Ellipse, semiaxis $a$}~~& & & &\\
{\scriptsize semiaxis $b \ge a$}            &~~ {\scriptsize $\pi
a b$}~~&~~{\scriptsize $4b E(\sqrt{1-\frac{a^2}{b^2}}\;)$}~~&~~
$\frac{2b}{\pi} {\scriptstyle E(\sqrt{1-\frac{a^2}{b^2}}\;)}$~~&~~
${\scriptstyle 2}+\frac{8b} {\pi^2 a} {\scriptstyle
E^2(\sqrt{1-\frac{a^2}{b^2}}\;)}$
\end{tabular}
\end{center}
\label{vol-2}
\end{table}

\subsection{Three Dimensions}

Table \ref{vol-3} provides volumes, surface areas, radii of mean curvature, and exclusion volumes
for some three-dimensional convex particles. This includes the sphere, Platonic solids, 
rectangular parallelepiped, spherocylinder and spheroid.  The radius of
mean curvature $\bar R$ can be obtained from the volume of the parallel body associated with
the convex body at distance $\epsilon$ that is of order $\epsilon^2$ and use
of relation (\ref{m-w-1}). The corresponding exclusion volume can then be found
employing relation (\ref{vex}) for $d=3$.

It is easy to extract $\bar R$ and $v_{\mbox{\scriptsize ex}}/v$ from the spheroid
results in Table \ref{vol-3} the ``prolate-needle" ($b/a \rightarrow \infty$) and ``oblate-needle" ($b/a \rightarrow 0$) limits,
which are respectively given by
\begin{equation}
{\bar R}\sim \frac{b}{2}, \qquad \frac{v_{\mbox{\scriptsize ex}}}{v} \sim \frac{3\pi}{4}\frac{b}{a}, \qquad b/a \rightarrow \infty,
\end{equation}
and
\begin{equation}
{\bar R}\sim \frac{\pi}{4}a, \qquad \frac{v_{\mbox{\scriptsize ex}}}{v} \sim \frac{3\pi}{4}\frac{a}{b}, \qquad b/a \rightarrow 0.
\end{equation}

\begin{table}[h]
\caption{Volumes, surface areas, radii of mean curvature, and exclusion volumes
for some convex particles $K$ in $\mathbb{R}^3$. The radius of mean curvature and exclusion volume correspond to randomly oriented convex
particles  having nonzero volumes.}
\begin{tabular}{c|c|c|c|c }
$K$& $v$ & $s$ & ${\bar R}$ & $v_{\mbox{\scriptsize ex}}/v$   \\
\hline
{\scriptsize Sphere, radius $a$} ~&~  $\frac{4\pi a^3}{3}$  ~&~{\scriptsize $4\pi a^2$} ~&~{\scriptsize $a~$} &~{\scriptsize $8$}    \\
{\scriptsize Icosahedron, side  $a$} ~&~ {\scriptsize$\frac{5}{12}(3+\sqrt{5}) a^3$} ~&~ {\scriptsize $5\sqrt{3} a^2$} ~&~ $\frac{30}{8\pi}{\scriptstyle \cos^{-1}(\frac{\sqrt{5}}{3})a}$ ~&~{\scriptsize 2}+  $\frac{90 \sqrt{3} \cos^{-1}(\frac{\sqrt{5}}{3})}{\pi (3+\sqrt{5})}$ \\
{\scriptsize Dodecahedron, side  $a$} ~&~ $\frac{15+7\sqrt{5}}{4} {\scriptstyle a^3}$ ~&~ {\scriptsize $3\sqrt{25+10\sqrt{5}}a^2$} ~&~ $\frac{30}{8\pi}{\scriptstyle \cos^{-1}(\frac{1}{\sqrt{5}})a}$ ~& \, {\scriptsize 2}+ $\frac{90 \sqrt{25+10\sqrt{5}} \cos^{-1}(\frac{1}{\sqrt{5}})}{\pi (15+7\sqrt{5})}$ \\
{\scriptsize Cube, side $a$} ~&  ~{\scriptsize $a^3$}  ~&~{\scriptsize $6a^2$} ~&~$\frac{3a}{4}$~ &~{\scriptsize $11$ }    \\
{\scriptsize Octahedron, side $a$} &~ {\scriptsize
$\frac{\sqrt{2}}{3} a^3$ }~&~ {\scriptsize $2\sqrt{3} a^2$} ~&~
$\frac{3}{2\pi}{\scriptstyle \cos^{-1}(\frac{1}{3})a}$ ~&~
{\scriptsize 2}+ $\frac{9}{\pi}
\sqrt{\frac{3}{2}}  {\scriptstyle \cos^{-1}(\frac{1}{3})}$\\
{\scriptsize Tetrahedron, side $a$} ~& ~{\scriptsize
$\frac{\sqrt{2}}{12} a^3$}~&~ {\scriptsize $\sqrt{3} a^2$}~&~
$\frac{3}{4\pi}{\scriptstyle \cos^{-1}(-\frac{1}{3})a}$~&
~{\scriptsize 2}+ $\frac{18}{\pi}
\sqrt{\frac{3}{2}}  {\scriptstyle \cos^{-1}(-\frac{1}{3})}$\\
\hspace{-0.3in} {\scriptsize Rectangular }~& & & &\\
{\scriptsize parallelpiped} ~& & & &\\
{\scriptsize  sides $a_1$, $a_2$, $a_3$}~&  ~${\scriptstyle  a_1
a_2 a_3}$  ~&~$ \sum_{i< j}^3 {\scriptstyle 2a_ia_j}$
~&~$\frac{1}{4} \sum_{i=1}^3 {\scriptstyle a_i}$~
&~{\scriptsize 2}+ $\frac{({ \sum_{i< j}^3} a_ia_j)({ \sum_{i=1}^3} a_i)} {a_1a_2a_3}$ \\
{\scriptsize Cylinder, radius $a$}~~& & & &\\
{\scriptsize height $h$}            &~ {\scriptsize $\pi a^2 h$}~&~{\scriptsize $2\pi a(a+h)$}~&~ $\frac{\pi a +h}{4}$~& {\scriptsize 2}+$\frac{(a+h)(\pi a+h)}{2 a h}$\\
{\scriptsize Spherocylinder, radius $a$}& & & &\\
{\scriptsize cylindrical height $h$ } &~ $\frac{\pi a^2(4a + 3h)}{3}$~&~{\scriptsize $2 \pi a(2a +h)$}~&~ $\frac{h+4a}{4}$~&~ {\scriptsize 2}+$\frac{3(2a+h)(4a+h)}{ a(4 a+3h)}$\\
{\scriptsize Prolate spheroid }~& & & &\\
{\scriptsize  semiaxes $a=c$, $b\ge a$ }~& & & &\\
{\scriptsize $e^2=1-(a/b)^2$ }~&  ~$\frac{4\pi a^2b}{3}$   ~&~
${\scriptstyle 2\pi a^2}(1+\frac{b}{a e}{\scriptstyle
\sin^{-1}e})$~ &~$\frac{b}{2}
(1+\frac{a^2}{b^2 e}{\scriptstyle \tanh^{-1}e)}$ ~&   {\scriptsize 2}+ $\frac{3}{2}  (1+\frac{b}{a e}{\scriptstyle \sin^{-1}e})$ \\
~& & & &\qquad$\times(1+\frac{a^2}{b^2 e}{\scriptstyle \tanh^{-1}}e)$\\
{\scriptsize Oblate spheroid} ~& & & &\\
{\scriptsize  semiaxes $a=c$, $b \le a$} ~& & & &\\
{\scriptsize $e^2=1-(b/a)^2$} ~&  ~$\frac{4\pi a^2b}{3}$   ~&~
${\scriptstyle 2\pi a^2}(1+\frac{b^2}{a^2 e} {\scriptstyle
\tanh^{-1}e})$~ &~$\frac{b}{2}(1+\frac{a}{b e}{\scriptstyle
\sin^{-1}e})
$ ~&    {\scriptsize 2}+ $\frac{3}{2}  (1+\frac{b^2}{a^2 e} {\scriptstyle \tanh^{-1}e})$ \\
~& & & &\qquad$\times(1+\frac{a}{b e}{\scriptstyle \sin^{-1}e})$\\
\end{tabular}
\label{vol-3}
\end{table}

Table \ref{vol-3-0} provides surface areas, radii of mean curvature, and exclusion volumes
for some three-dimensional convex plates. This includes the circular, square, 
equilateral triangular, rectangular and elliptical plates.
The radius of
mean curvature $\bar R$ can be easily obtained from the volume of the parallel body associated with
the convex body at distance $\epsilon$ that is of order $\epsilon^2$ and use
of relation (\ref{m-w-1}). The corresponding exclusion volume can then be found
using relation (\ref{vex}) for $d=3$. Alternatively, in the cases of
circular, square and rectangular plates, $\bar R$ and $v_{\mbox{\scriptsize ex}}$ 
can be obtained from Table \ref{vol-3} as special cases of the cylinder (when $h=0$)
and rectangular parallelpiped ($a_1=a_2=a$, $a_3=0$ for the square plate
and $a_3=0$ for the rectangular plate). To get $v_{\mbox{\scriptsize ex}}$
in these cases, one must first multiply $v_{\mbox{\scriptsize ex}}/v$   in Table \ref{vol-3} by the volume $v$ and then take
the aforementioned limits.

\begin{table}[!htp]
\caption{Surface areas, radius of mean curvature, and exclusion volumes
for some convex plates $K$ in $\mathbb{R}^3$ possessing a volume $v=0$. The 
radius of mean curvature and exclusion volume correspond to randomly oriented convex
plates.}
\begin{tabular}{c|c|c|c}
$K$&  $s$ & ${\bar R}$ & $v_{\mbox{\scriptsize ex}}$   \\
\hline
Circular plate, radius $a$ ~&~$2\pi a^2$ ~&~$\frac{\pi}{4}a$ ~ &~$\pi^2 a^3$    \\
Square plate, side $a$ ~&~$2 a^2$ ~&~$\frac{1}{2}a$ ~ &~$2 a^3$    \\
Equilateral triangular plate, side $a$ ~&~$\frac{\sqrt{3}}{2} a^2$ ~&~$\frac{3}{8}a$ ~ &~$\frac{3\sqrt{3}}{8} a^3$    \\
Rectangular plate, sides $a_1$, $a_2$ ~&~$2a_1a_2$ ~&~$\frac{1}{4}(a_1+a_2)$ ~ &~$a_1a_2(a_1+a_2)$    \\
Elliptical plate, semiaxis $a$~~& & & \\
semiaxis $b \ge a$    ~& $2\pi ab$  ~&~ $\frac{1}{2}b
E(\sqrt{1-a/b}\;)$~&~ $2\pi a b^2 E(\sqrt{1-a/b}\;)$
\end{tabular}
\label{vol-3-0}
\end{table}

\section{Radii of Mean Curvature and Exclusion Volumes for Various Hyperparticles}
\label{d}

We  present here explicit formulas for the radii of mean curvature and exclusion
volumes for a variety of hyperparticles in $\mathbb{R}^d$ with random orientations possessing 
both nonzero and zero volumes (hyperplates).

\subsection{Hypercube and Hyperrectangular Parallelpiped}

\subsubsection{Hypercube}

The determination of the radius of mean curvature ${\bar R}$ for a $d$-cube (hypercube)
of edge (or side) length $a$ is easily obtained from the $d$-dimensional generalization of Steiner's formula. Associated
with the $\epsilon$ neighborhood of an edge  is a portion of a $(d-1)$-dimensional hypercylinder
of height $a$ and radius $\epsilon$,  the volume of which is given by
\begin{equation}
v_{\mbox{\scriptsize cyl}}= v_s(d-1;\epsilon) \;a=\frac{\pi^{(d-1)/2}
}{\Gamma(1+\frac{d-1}{2})} \;a \,\epsilon^{d-1},
\end{equation}
where $v_s(d;a)$ is the volume of a $d$-dimensional hypersphere of radius $a$ explicitly given by (\ref{sphere}). Hence, the volume associated with a hypercube edge is $v_{\mbox{\scriptsize cyl}}/2^{d-1}$.
Since a hypercube has a total of $d 2^{d-1}$ edges,  the total volume associated
with all of the edges is $d v_{\mbox{\scriptsize cyl}}$. Therefore, equating this volume with the term $W_{d-1} \epsilon^{d-1}$
in Steiner's formula (\ref{Stein-d}) and using (\ref{m-w-1}) yields
\begin{equation}
\Omega(d) {\bar R} \epsilon^{d-1} = d v_{\mbox{\scriptsize cyl}},
\end{equation}
and hence the radius of mean curvature for a hypercube of side length $a$ is given by
\begin{equation}
{\bar R} =\frac{\Gamma(1+\frac{d}{2})}{\sqrt{\pi} \Gamma(\frac{d+1}{2})}\;a.
\label{R-cube}
\end{equation}
For large $d$, we obtain the asymptotic result
\begin{equation}
{\bar R} \sim  \sqrt{\frac{d}{2\pi}},  \qquad d \rightarrow \infty.
\label{R-cube-asym}
\end{equation}
Not surprisingly, for large $d$,  the longest diagonal, which grows like $\sqrt{d}$,
is the dominant contribution to ${\bar R}$.

Using Eqs. (\ref{vex}) and (\ref{R-cube}) and the fact that the surface area of a hypercube of side length $a$
are given by $s=2d \;a^{d-1}$ and $v=a^d$, respectively, we obtain the following expression
for the exclusion volume:
\begin{equation}
\frac{v_{\mbox{\scriptsize ex}}}{v}=2 +\frac{4(2^{d-1}-1)}{\sqrt{\pi}} \frac{\Gamma(1+\frac{d}{2})}{\Gamma(\frac{d+1}{2})}.
\label{vex-cube}
\end{equation}
In the large-$d$ limit, this expression and  (\ref{R-cube-asym}) yield the asymptotic result
\begin{equation}
\frac{v_{\mbox{\scriptsize ex}}}{v} \sim \sqrt{\frac{2 d}{\pi}}\;2^d, \qquad d \rightarrow \infty.
\label{vex-cube-asym}
\end{equation}

\subsubsection{Hyperrectangular Parallelpiped}

Consider a $d$-rectangular parallelpiped (hyperrectangular parallelpiped) with edges
of side lengths $a_1$, $a_2$, $\cdots$, $a_d$. Using a similar analysis as for
the hypercube, we find the radius of mean curvature is given by
\begin{equation}
{\bar R} =\frac{\Gamma(1+\frac{d}{2})}{d\sqrt{\pi} \;\Gamma(\frac{d+1}{2})}\;\sum_{i=1}^d a_i.
\label{R-rect}
\end{equation}
Employing Eqs. (\ref{vex}) and (\ref{R-rect}) and the fact that the surface area of a hyperrectangular parallelpiped
of side lengths $a_1,a_2,\ldots,a_d$
is given by $s=\sum_{i<j<k<l\cdots}^d 2a_ia_ja_k a_l\cdots $ (where 
$a_ia_ja_k a_l \cdots$ is the product of $d-1$ different side lengths) and $v=\prod_{i=1}^d a_i$, we obtain the following expression
for the exclusion volume:
\begin{equation}
\frac{v_{\mbox{\scriptsize ex}}}{v}=2 +\frac{2(2^{d-1}-1)}{d^2\sqrt{\pi}} \frac{\Gamma(1+\frac{d}{2})}{\Gamma(\frac{d+1}{2})}
\frac{\displaystyle \left(\sum_{i<j<k<l\cdots }^d 2a_ia_ja_k a_l\cdots\right)\left(\sum_{i=1}^d a_i\right)}{\prod_{i=1}^d a_i}.
\label{vex-rect}
\end{equation}

Consider the special case in which there are $d -1$ edges with side length $a$
and the remaining edge with side length $b$. Then, formulas (\ref{R-rect})
and (\ref{vex-rect}) simplify as follows:
\begin{equation}
{\bar R} =\frac{\Gamma(1+\frac{d}{2})}{\sqrt{\pi} d\;\Gamma(\frac{d+1}{2})}[(d-1)a+b]\;,
\label{R-rect-2}
\end{equation}
\begin{equation}
\frac{v_{\mbox{\scriptsize ex}}}{v}=2 +\frac{4(2^{d-1}-1)}{d^2\sqrt{\pi}} \frac{\Gamma(1+\frac{d}{2})}{\Gamma(\frac{d+1}{2})}
\left[(d-1)\frac{b}{a} +1\right] \left[(d-1)\frac{a}{b} +1\right].
\label{vex-rect-2}
\end{equation}
We shall subsequently make use of these relations in deriving certain hyperplate results.

\subsection{Hyperspherocylinder}

Consider a $d$-spherocylinder (hyperspherocylinder) of height $h$ and radius $a$,
which can be decomposed into hypercylinder of height $h$ and two hemispherical
caps of radius $a$ (i.e., hypersphere of radius $a$). The volume of order $\epsilon^{d-1}$ associated
with the $\epsilon$ neighborhood of a hyperspherocylinder is
\begin{displaymath}
\frac{\pi^{(d-1)/2} \epsilon^{d-1} h}{\Gamma(\frac{d+1}{2})} + \frac{\pi^{d/2} d \epsilon^{d-1} a}{\Gamma(1+d/2)},
\end{displaymath}
which when equated with the term $W_{d-1} \epsilon^{d-1}$ in Steiner's formula 
(\ref{Stein-d}) and using (\ref{m-w-1})  gives
\begin{equation}
\Omega(d) {\bar R} \epsilon^{d-1}=  \frac{d\pi^{(d-1)/2} \epsilon^{d-1} h}{\Gamma(\frac{d+1}{2})} + \frac{\pi^{d/2} d \epsilon^{d-1} a}{\Gamma(1+d/2)},
\end{equation}
and hence the radius of mean curvature for a hyperspherocylinder is given by
\begin{equation}
{\bar R} =\frac{\Gamma(1+\frac{d}{2}) }{d\sqrt{\pi}\; \Gamma(\frac{d+1}{2})}\;h +a.
\label{R-sphero}
\end{equation}
For large $d$, we find the asymptotic result
\begin{equation}
{\bar R} \sim  \sqrt{\frac{1}{2\pi d}}\;h +a,  \qquad d \rightarrow \infty.
\label{R-sphero-asy}
\end{equation}
We see that if $h$ grows mores slowly with dimension than $\sqrt{d}$,
\begin{equation}
{\bar R} \sim  a, \qquad d \rightarrow \infty.
\end{equation}

The surface area and volume of a spherocylinder are given respectively by
\begin{equation}
s= s_s(d-1;a) h + s_s(d;a)
\end{equation}
and
\begin{equation}
v= v_s(d-1;a) h +v_s(d;a),
\end{equation}
where $s_s(d;a)$ and $v_s(d;a)$ are the surface area of a $d$-dimensional hypersphere of radius $a$ given
by (\ref{s_s}) and (\ref{sphere}), respectively.
These relations combined with Eqs. (\ref{vex}) and (\ref{R-sphero}) yield the exclusion
volume to be given by
\begin{equation}
\frac{v_{\mbox{\scriptsize ex}}}{v}=2 +\frac{2(2^{d-1}-1)}{d}
\left[\frac{s_s(d-1;a) h + s_s(d;a)}{v_s(d-1;a) h +v_s(d;a)} \right]\left[\frac{\Gamma(1+\frac{d}{2}) h}{d\sqrt{\pi}
\;\Gamma(\frac{d+1}{2})} +a\right].
\end{equation}
In the large-$d$ limit, this relation yields the asymptotic result
\begin{equation}
\frac{v_{\mbox{\scriptsize ex}}}{v} \sim \left[\frac{h}{a} \sqrt{\frac{1}{2\pi d}}+ 1\right]  2^d,
\label{vex-asy}
\end{equation}
where we have used (\ref{R-sphero-asy}). We observe that in order for the contribution to the the scaled
exclusion volume from the cylindrical portion, first term in (\ref{vex-asy}), not to vanish  as $d$ becomes large,
the aspect ratio $h/a$ must grow
as fast as $\sqrt{d}$ or faster.

\subsection{Hyperoctahedron and Hypertetrahedron}

A regular $d$-crosspolytope or regular hyperoctahedron (also known
as the orthoplex) is the $d$-dimensional generalization
of the three-dimensional regular octahedron. For $d=1$ and $d=2$, the regular hyperoctahedron is a line
and square, respectively. A regular $d$-simplex or hypertetrahedron is the $d$-dimensional
generalization of the three-dimensional regular tetrahedron. For $d=1$ and $d=2$, the 
regular hypertetrahedron is a line
and equilateral triangle, respectively. For $d\ge 5$, there are only three types of convex regular polytopes: the regular hypercube, hyperoctahedron and hypertetrahedron \cite{Co73}.
For a regular hyperoctahedron and regular hypertetrahedron, the radius of mean curvature
is not as easy to derive from Steiner's formula, but integral formulas for it
have been derived \cite{Ha79,Bet93,Hen97}.

\subsubsection{Regular Hyperoctahedron ($d$-Crosspolytope) }

For a regular  hyperoctahedron  with unit side length,
the radius of mean curvature is given by \cite{Bet93,Hen97}
\begin{equation}
{\bar R} =\frac{\sqrt{2}\;d(d-1)\; \Gamma( \frac{d}{2})}{2\sqrt{2} \pi \Gamma(\frac{d+1}{2})}\; I_O(d),
\label{R-oct}
\end{equation}
where
\begin{equation}
I_O(d) = \int_{0}^{\infty}  \exp(-2x^2) \; \mbox{erf}(x)^{d-2}\; dx.
\end{equation}
Note that for $d=2$ and $d=3$, we respectively have
\begin{equation}
I_O(2)=\frac{\sqrt{\pi}}{2\sqrt{2}}
\end{equation}
and
\begin{equation}
I_O(3)=\frac{1}{\sqrt{2\pi}} \tan^{-1}(\frac{1}{\sqrt{2}}).
\end{equation}
Substitution of these formulas into (\ref{R-oct}) recovers
the results for $\bar R$ listed in Tables \ref{vol-2} and \ref{vol-3}
for the square and octahedron, respectively.
For large $d$, it is straightforward to show that
\begin{equation}
I_O(d) \sim \frac{\pi \sqrt{\ln d}}{2d^2},
\end{equation}
which implies
\begin{equation}
{\bar R} \sim \sqrt{\frac{\ln d}{d}}, \qquad d \rightarrow \infty.
\label{R-oct-asy}
\end{equation}

The surface area and volume of a  regular hyperoctahedron are given respectively by
\begin{equation}
s= \frac{2^{(d+1)/2}\sqrt{d}}{(d-1)!}
\end{equation}
and
\begin{equation}
v=  \frac{2^{d/2}}{d!}.
\end{equation}
These relations combined with Eqs. (\ref{vex}) and (\ref{R-oct}) yield the exclusion
volume to be given by
\begin{equation}
\frac{v_{\mbox{\scriptsize ex}}}{v}= 2+\frac{4 d^{3/2} (d-1)\;\Gamma(\frac{d}{2})(2^{d-1}-1)}{\pi\Gamma(\frac{d-1}{2})}I_O(d).
\end{equation}
In the large-$d$ limit, this relation and (\ref{R-oct-asy}) give the asymptotic result
\begin{equation}
\frac{v_{\mbox{\scriptsize ex}}}{v} \sim \sqrt{2\ln(d)} \; 2^d,  \qquad d \rightarrow \infty.
\end{equation}

\subsubsection{Regular Hypertetrahedron ($d$-Simplex)}

For a regular  hypertetrahedron with unit side length, the radius of mean curvature is given by \cite{Ha79,Hen97}
\begin{equation}
{\bar R} =\frac{d(d+1)\; \Gamma(\frac{d}{2})}{2\sqrt{2}\;\pi \Gamma(\frac{1+d}{2})}\; I_T(d),
\label{R-tetra}
\end{equation}
where
\begin{equation}
I_T(d) = \int_{-\infty}^{\infty}  \exp(-2x^2) \left[ \frac{1+\mbox{erf}(x)}{2} \right]^{d-1}\; dx.
\end{equation}
Note that for $d=2$ and $d=3$, we respectively have
\begin{equation}
I_T(2)=\frac{\pi}{2\sqrt{2}}
\end{equation}
and
\begin{equation}
I_T(3)=\frac{1}{2\sqrt{2\pi}} \cos^{-1}(-\frac{1}{3}).
\end{equation}
Substitution of these formulas into (\ref{R-tetra}) recovers
the results for $\bar R$ listed in Tables \ref{vol-2} and \ref{vol-3}
for the equilateral triangle and tetrahedron, respectively.
For large $d$, it is simple to show that
\begin{equation}
I_T(d) \sim \frac{2\pi \sqrt{\ln d}}{d^2},
\end{equation}
which implies
\begin{equation}
{\bar R} \sim \sqrt{\frac{\ln d}{d}}, \qquad d \rightarrow \infty.
\label{R-tetra-asy}
\end{equation}
We see that in the large-$d$ limit, the radii of mean curvature for a regular hyperoctahedron
and regular hypertetrahedron  become identical.

The surface area and volume of a  regular hypertetrahedron are given respectively by
\begin{equation}
s= \frac{\sqrt{d}(d+1)}{2^{(d-1)/2} (d-1)!}
\end{equation}
and
\begin{equation}
v=  \frac{\sqrt{d+1}}{2^{d/2} d!}.
\end{equation}
These relations combined with Eqs. (\ref{vex}) and (\ref{R-tetra}) yield the exclusion
volume to be given by
\begin{equation}
\frac{v_{\mbox{\scriptsize ex}}}{v}=
2+\frac{d^{3/2}(d+1)^{3/2}\;\Gamma(\frac{d}{2})(2^{d-1}-1)}{\pi\Gamma(\frac{d-1}{2})}I_T(d).
\end{equation}
In the large-$d$ limit, this relation and  (\ref{R-tetra-asy}) yield the asymptotic result
\begin{equation}
\frac{v_{\mbox{\scriptsize ex}}}{v} \sim \sqrt{2d\ln(d)} \; 2^d,  \qquad d \rightarrow \infty.
\end{equation}

\begin{figure}
\begin{center}
\includegraphics[height=7.5cm,keepaspectratio,clip=]{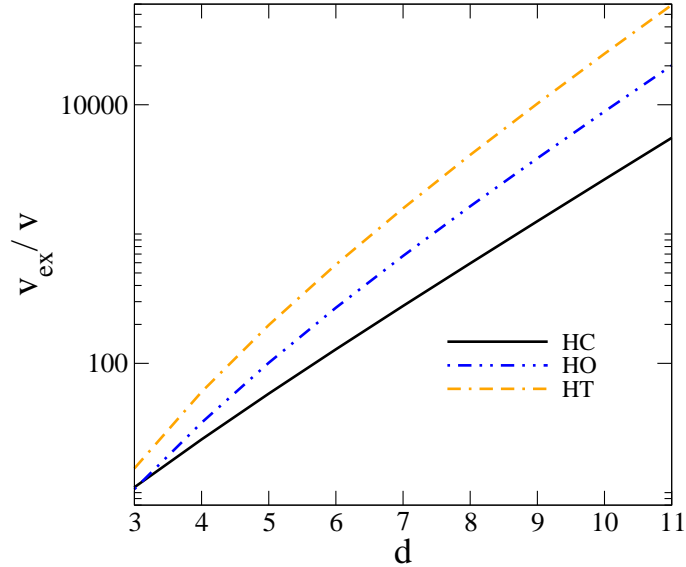}
\end{center}
\caption{(Color online) The dimensionless exclusion
volume $v_{\mbox{\scriptsize ex}}/v$ versus dimension $d$ 
for the three convex regular polytopes:  hypercube, hyperoctahedron and hypertetrahedron.}
\label{poly}
\end{figure}

Among the three possible convex regular polytopes in high dimensions (hypercube, hyperoctahedron and hypertetrahedron), the hypertetrahedron has the largest  dimensionless exclusion
volume $v_{\mbox{\scriptsize ex}}/v$ for any fixed dimension $d\ge 3$, while
the hypercube has the smallest value. 
Figure \ref{poly} shows $v_{\mbox{\scriptsize ex}}/v$ versus $d$ for $ 3 \le d \le 11$
for the three convex regular polytopes.

\subsection{Hyperplates}

\subsubsection{$(d-1)$-cube in $\mathbb{R}^d$}

Consider a hyperplate in $d$-dimensional space that is a $(d-1)$-cube
with side length $a$. The radius of mean curvature and exclusion volume
of such a hyperplate are trivially obtained from special cases of relations (\ref{R-rect-2}) and (\ref{vex-rect-2}). That is, setting $b=0$ in (\ref{R-rect-2}) yields 
\begin{equation}
{\bar R} =\frac{(d-1)\Gamma(1+\frac{d}{2})}{d\sqrt{\pi} \;\Gamma(\frac{d+1}{2})}\;a\;.
\label{R-cube-2}
\end{equation}
The corresponding exclusion volume is extracted from (\ref{vex-rect-2}) by first
multiplying both sides of the equation by the volume $v=a^{d-1}b$, and then
setting $b=0$, yielding
\begin{equation}
v_{\mbox{\scriptsize ex}}= \frac{4(d-1)(2^{d-1}-1)\Gamma(1+\frac{d}{2})}{d^2\sqrt{\pi}\;\Gamma(\frac{d+1}{2})}\;a^d,
\label{vex-cube-2}
\end{equation}
which, not surprisingly, is smaller than the exclusion volume (\ref{vex-cube}) for a $d$-cube in $\mathbb{R}^d$ at fixed $d$.
Not surprisingly, in the high-$d$ limit, formula (\ref{R-cube-2}) for the radius of mean curvature
tends to the same asymptotic expression (\ref{R-cube-asym})
for a $d$-cube in $\mathbb{R}^d$. This is not the case for the high-$d$ limit
of (\ref{vex-cube-2}) for the exclusion volume, which is given by
\begin{equation}
v_{\mbox{\scriptsize ex}} \sim \sqrt{\frac{2 }{d\pi}}\;2^d \;a^d,  \qquad d \rightarrow \infty,
\end{equation}
and should be  compared to (\ref{vex-cube-asym}), which grows faster as $d$ increases.

Note that in the case $d=3$, the formulas (\ref{R-cube-2}) and (\ref{vex-cube-2}) give the same results reported
in Table \ref{vol-3} for the instance of a square plate in $\mathbb{R}^3$. 
For $d=2$, they also yield the relations (\ref{line-1}) and (\ref{line-2}) for the
case of a randomly oriented line in $\mathbb{R}^2$.

\subsubsection{$(d-1)$-sphere in $\mathbb{R}^d$}

Consider a hyperplate in $d$-dimensional space that is a $(d-1)$-sphere
of radius $a$. For such a hyperplate, $\bar R$ 
can be obtained from the corresponding relation (\ref{R-cube-2}) for a cubical
hyperplate by multiplying
the latter by the factor
\begin{displaymath}
\frac{\sqrt{\pi}\;\Gamma(\frac{d}{2})}{\Gamma(\frac{d+1}{2})},
\end{displaymath}
which is the ratio of radius of mean curvature ${\bar R}=a$ of a $(d-1)$-sphere 
in $\mathbb{R}^{d-1}$ to that of the $(d-1)$-cube of side length $a$ in $\mathbb{R}^{d-1}$, the latter of  which is obtained by 
using (\ref{R-cube}). Thus, the  radius of mean
of curvature for a $(d-1)$-sphere of radius in $\mathbb{R}^{d}$ is given by
\begin{equation}
{\bar R} =\frac{(d-1)\;\Gamma(\frac{d}{2})^2}{2\;\Gamma(\frac{d+1}{2})^2}\;a\;.
\label{R-sphere-2}
\end{equation}
To illustrate the fact that this simple mapping 
is equivalent to the explicit calculation of $\bar R$ for a $(d-1)$-sphere, we carry
out such a computation in Appendix A for a 3-sphere in $\mathbb{R}^4$. 
Using the fact that the surface area of a $(d-1)$-sphere in $\mathbb{R}^d$
is $s=2v_s(d-1;a)$, where  $v_s(d;a)$ is  given by (\ref{sphere}), we obtain
the $d$-dimensional exclusion volume to be
\begin{equation}
v_{\mbox{\scriptsize ex}}= \frac{2(d-1)(2^{d-1}-1)\pi^{(d-1)/2}\Gamma(\frac{d}{2})^2}{d \;\Gamma(\frac{d+1}{2})^3}\;a^{d} < \frac{2^d\pi^{d/2}}{\Gamma(1+d/2)}\;a^d,
\label{vex-sphere-2}
\end{equation}
where the upper bound is the exclusion volume of a $d$-sphere in $\mathbb{R}^d$ \cite{To12a}.
Not surprisingly, in the high-$d$ limit, formula (\ref{R-sphere-2}) for the radius of mean curvature
tends to the radius $a$.  In this asymptotic limit, relation (\ref{vex-sphere-2}) becomes
\begin{equation}
v_{\mbox{\scriptsize ex}} \sim \frac{2^{3/2}}{d\;\sqrt{\pi}}\left(\frac{2\pi \exp(1)}{d}\right)^{d/2} \;2^d\,a^{d} < \frac{1}{\sqrt{d\pi}}\left(\frac{2\pi \exp(1)}{d}\right)^{d/2} \;2^d\;a^d.
\end{equation}

Observe that in the case $d=3$, the formulas (\ref{R-sphere-2}) and (\ref{vex-sphere-2}) give the same results reported
in Table \ref{vol-3} for the instance of a circular plate in $\mathbb{R}^3$. 
For $d=2$, they also yield the relations (\ref{line-1}) and (\ref{line-2}) for the
case of a randomly oriented line in $\mathbb{R}^2$.

\section{Conjecture for the Maximum-Threshold Hyperparticle 
Shape Among All Convex Bodies in $\mathbb{R}^d$}
\label{CONJ}


Recall that the dimensionless exclusion volume $v_{\mbox{\scriptsize ex}}/v$,
among all convex bodies in $\mathbb{R}^d$ with a nonzero  $d$-dimensional volume,
is minimized for hyperspheres [see Sec. \ref{explicit}] and its
threshold $\eta_c$ exactly tends to $v/v_{\mbox{\scriptsize ex}}=2^{-d}$ in
the high-dimensional limit [cf. (\ref{asy})]. These properties
together with the principle that
low-$d$ percolation properties encode high-$d$ information
\cite{To12a}, leads us to the following conjecture:

{\sl Conjecture: The percolation threshold $\eta_c$ among all
systems of overlapping randomly oriented convex hyperparticles in $\mathbb{R}^d$ having nonzero
volume is maximized
by that for hyperspheres, i.e.,
\begin{equation}
(\eta_c)_S \ge \eta_c,
\label{conj}
\end{equation}
where $(\eta_c)_S$ is the threshold of overlapping hyperspheres.}

We note that similar reasoning leading to the aforementioned
conjecture (\ref{conj}) for convex hyperparticles of nonzero volume also suggests that the dimensionless exclusion volume
 $v_{\mbox{\scriptsize ex}}/v_{\mbox{\scriptsize eff}}$ associated with a convex ($d-1$)-dimensional hyperplate in $\mathbb{R}^d$
is minimized by the ($d-1$)-dimensional hypersphere, which
consequently would have the highest percolation threshold among all convex hyperplates.
Recall that $v_{\mbox{\scriptsize eff}}$ the effective $d$-dimensional volume
associated with the zero-volume hyperplate of interest, defined
by  (\ref{plate_vol}).

\section{Accurate Scaling Relation for the Percolation Threshold of Overlapping Convex Hyperparticles}
\label{approx}

Recall that  lower bound  (\ref{aniso}) on
the threshold $\eta_c$ of overlapping convex hyperparticles
of general shapes with nonzero volumes and  a specified orientational distribution is 
determined by the dimensionless exclusion volume  $v_{\mbox{\scriptsize ex}}/v$, which also provides
the exact high-$d$ asymptotic limit [cf. (\ref{aniso_bd})]. A special case
of this lower bound is the inequality (\ref{iso}), which
is valid for hyperspheres as well as aligned centrally symmetric
particles for which $v_{\mbox{\scriptsize ex}}/v=2^d$.



Guided by the aforementioned high-dimensional behavior of $\eta_c$, conjecture (\ref{conj})
for hyperspheres, and the functional form of the lower bounds
(\ref{isot}) and (\ref{aniso}), we propose the following scaling
law for the threshold $\eta_c$ of overlapping nonspherical convex
hyperparticles of arbitrary shape and
orientational distribution having nonzero volumes for any dimension $d$:
\begin{eqnarray}
\eta_c &\approx& \left(\frac{v_{\mbox{\scriptsize ex}}}{v}\right)_S \left(\frac{v}{v_{\mbox{\scriptsize ex}}}\right)(\eta_c)_S\nonumber\\
&=& 2^d \left(\frac{v}{v_{\mbox{\scriptsize ex}}}\right)(\eta_c)_S,
\label{scaling}
\end{eqnarray}
where $(\eta_c)_S$ and $(v_{\mbox{\scriptsize ex}}/v)_S=2^d$ are
the percolation threshold and dimensionless exclusion volume,
respectively, for hyperspheres in dimension $d$. Thus, given the
percolation threshold for the {\it reference system} of hyperspheres $(\eta_c)_S$ and the
dimensionless exclusion volume $v_{\mbox{\scriptsize ex}}/v$ for
some general nonspherical convex hyperparticle shape with a
specified orientational distribution, one can estimate 
the threshold $\eta_c$ of such a system of overlapping
hyperparticles across dimensions via the scaling relation (\ref{scaling}).
Note that this scaling relation becomes exact in the high-$d$ limit.
In the next section, we will show that this scaling relation
provides reasonably accurate estimates of $\eta_c$ in low dimensions
and, hence, must become increasingly accurate as $d$ becomes large.
This is yet another manifestation of the principle that
low-dimensional percolation properties encode high-dimensional information \cite{To12a}.

It is noteworthy that the scaling relation (\ref{scaling}) not only provides
a good approximation for the threshold $\eta_c$ of overlapping
nonspherical hyperparticles, it yields an upper bound on $\eta_c$,
and hence a relatively tight one. Explicitly, we have
the following $d$-dimensional inequality:
\begin{eqnarray}
\eta_c \ge 2^d \left(\frac{v}{v_{\mbox{\scriptsize ex}}}\right)(\eta_c)_S.
\label{upper}
\end{eqnarray}
This bounding
property is a consequence of an observation made in Ref. \onlinecite{To12a},
namely, that, at fixed $d$, the lower bound (\ref{aniso}) on $(\eta_c)_S$ for overlapping
hyperspheres converges more slowly to its exact asymptotic value
of $2^{-d}$ than does the lower bound (\ref{aniso}) for nonspherical
hyperparticles with a specified orientational distribution. It then immediately follows
from the functional form of  (\ref{scaling})
that it bounds $\eta_c$ for nonspherical hyperparticles
from above and converges to the exact
asymptotic value of $ v/v_{\mbox{\scriptsize ex}}$ 
in the large-$d$ limit [cf. (\ref{aniso_bd})].

For a zero-volume convex  ($d-1$)-dimensional hyperplate in $\mathbb{R}^d$, 
it is more appropriate to choose  the {\it reference system} 
to be ($d-1$)-dimensional hyperspheres of characteristic radius $r$, yielding
the scaling relation for the threshold for some other hyperplate shape to be
\begin{eqnarray}
\eta_c &\approx& \left(\frac{v_{\mbox{\scriptsize ex}}}{v_{\mbox{\scriptsize eff} }}\right)_{SHP} \left(\frac{v}{v_{\mbox{\scriptsize ex}}}\right)(\eta_c)_{SHP}\nonumber\\
&=& 2^d \left(\frac{v_{\mbox{\scriptsize eff} }}{v_{\mbox{\scriptsize ex}}}\right)(\eta_c)_{SHP}.
\label{plate-scaling}
\end{eqnarray}
Here $(\eta_c)_{SHP}$ is the percolation threshold for a $(d-1)$-dimensional hypersphere
and $v_{\mbox{\scriptsize eff} }$ is its effective volume, as defined by relation (\ref{plate_vol}).

We will test these scaling laws and bounds for a variety of hyperparticle
shapes across various dimensions using the exclusion volumes
presented in Secs. \ref{2-3} and \ref{d} and comparing these
results to available numerically-determined threshold estimates, including
those values that we have obtained below  for the Platonic solids.

\section{Application of the Scaling Relation and Bounds}
\label{appl}

Here, we apply the lower bound (\ref{aniso}) and scaling relation (\ref{scaling})
to estimate $\eta_c$ for a wide class of
randomly oriented overlapping nonspherical hyperparticles in $\mathbb{R}^d$, 
including, among other shapes, various polygons for $d=2$, Platonic solids, spherocylinders, parallepipeds
and zero-volume plates for $d=3$ and their appropriate generalizations for $d \ge 4$, when possible.
We  test these estimates against available numerical results 
for $\eta_c$ in $\mathbb{R}^2$ and $\mathbb{R}^3$, including
those obtained here for 
Platonic solids using the rescaled-particle method \cite{paper2}.

\subsection{Two Dimensions}

\begin{table}[!htp]
\caption{Percolation threshold $\eta_c$ of certain overlapping
convex particles $K$ with random orientations in $\mathbb{R}^2$, estimated
using Eq. (\ref{scaling}) and the associated threshold values
$\eta^*_c$ for squares \cite{randcube, cube2} and ellipses \cite{ellip1, ellip, ellip2} 
obtained from previous numerical simulations.
The threshold of overlapping circles \cite{ziff0, ziff1} is also given,
which is employed to compute the estimates.} 
\begin{tabular}{c|c|c}
$K$~ &  ~~$\eta^*_c$~~ & ~~$\eta_c$~~    \\
\hline
Circle ~ & ~~1.1281~~ & ~~~ \\
Square ~ & ~~0.9822~~ & ~~0.9925~ \\
Ellipse $b=2a$~ & ~~0.76 ~~ & ~~1.107~ \\
Ellipse $b=5a$~ & ~~0.607 ~~ & ~~0.7174~ \\
Ellipse $b=10a$~ & ~~0.358 ~~ & ~~0.4402~ \\
Ellipse $b=100a$~ & ~~0.0426 ~~ & ~~0.0543~ \\
Octagon ~ & ~~~~ & ~~1.0980~ \\
Hexagon ~ & ~~~~ & ~~1.0730~ \\
Pentagon ~ & ~~~~ & ~~1.0463~ \\
Equilateral Triangle ~ & ~~~~ & ~~0.8501~ \\
Rectangle $a_2=2a_1$~ & ~~~~ & ~~0.9275~ \\
Spherocylinder $h=2a$ ~ & ~~~~ & ~~1.0357~ \\
\end{tabular}
\label{tab_scaling_2d}
\end{table}

Table \ref{tab_scaling_2d} gives the percolation threshold
$\eta_c$ estimated using the scaling relation (\ref{scaling}) for
a variety of randomly oriented overlapping particles in
$\mathbb{R}^2$. In the cases of squares and ellipses with
different aspect ratios, the associated threshold value $\eta^*_c$
obtained from numerical simulations \cite{randcube, cube2, ellip1, ellip, ellip2}
are also given in the table. Clearly, for most
of randomly oriented nonspherical particles, the scaling relation
not only bounds $\eta_c$ from above, consistent with inequality
(\ref{upper}), the bounds are relatively tight. Since
the scaling relation already provides a good estimate
for $d=2$, it will increasingly become accurate
as $d$ increases for reasons discussed
in Sec. \ref{approx}. We will see below that this is indeed
the case for $d=3$.

\subsection{Three Dimensions}

\subsubsection{Nonzero-Volume Convex Bodies}

The percolation thresholds for randomly oriented
overlapping Platonic solids (i.e., tetrahedron, octahedron,
dodecahedron, icosahedron and cube) are estimated using a recently
developed rescaled-particle method. Implementation of this 
algorithm for the Platonic solids is described in 
Appendix B of this paper. In Ref. \onlinecite{paper2}, the rescaled-particle
method was applied to obtain accurate estimates of $\eta_c$ for
overlapping hyperspheres and oriented hypercubes.

\begin{table}[!htp]
\caption{Numerically estimated percolation threshold $\eta_c$ of randomly
oriented overlapping Platonic solids and the associated lower
bound value $\eta_L$ given by (\ref{aniso}). The numerical
values of the dimensionless exclusion volumes of these polyhedra
$v_{\mbox{\scriptsize ex}}/v$, whose analytical expressions are
given in Table \ref{vol-3}, are also included}
\begin{tabular}{c|@{\hspace{0.5cm}}c@{\hspace{0.5cm}}|@{\hspace{0.5cm}}c@{\hspace{0.5cm}}|@{\hspace{0.5cm}}c}
$K$&  $v_{\mbox{\scriptsize ex}}/v$ & $\eta_L$ & ${\eta_c}$   \\
\hline
Tetrahedron ~& ~15.40743~&~ 0.06493 ~&~ 0.1701$\pm$0.0007  \\
Cube ~& ~11~ &~ 0.09090  ~&~ 0.2443$\pm$0.0005     \\
Octahedron ~& ~10.63797~&~ 0.09398 ~&~ 0.2514$\pm$0.0006   \\
Dodecahedron ~& ~9.12101~&~ 0.1096 ~&~ 0.2949$\pm$0.0005    \\
Icosahedron ~& ~8.91526~&~ 0.1126 ~&~  0.3030$\pm$0.0005   \\
\end{tabular}
\label{tab_platonic}
\end{table}


Table \ref{tab_platonic} lists the numerical estimates  of
$\eta_c$, the associated lower-bound values $\eta_L$ given by (\ref{aniso}), and
the numerical values of the dimensionless exclusion volumes for
these polyhedra $v_{\mbox{\scriptsize ex}}/v$ (whose analytical
expressions are given in Table \ref{vol-3}). Observe that the tetrahedron
and icosahedron possess the smallest and largest percolation thresholds, respectively,
among the Platonic solids, which is confirmed in Table  \ref{tab_platonic}.
As discussed above, this implies that the the tetrahedron
and icosahedron possess the largest and smallest dimensionless exclusion volumes, respectively,
among the Platonic solids. We note that our estimate of $\eta_c$ for randomly oriented overlapping cubes is
consistent with that reported in Ref.~\onlinecite{randcube}, i.e.,
$\eta_c = 0.2444\pm0.0003$, which verifies the accuracy of our approach
for estimating $\eta_c$. The number within the parentheses
represents the error in the last digit. These results will be
employed below to verify the accuracy of the general scaling
relation (\ref{scaling}) for estimating $\eta_c$ of nonspherical
convex hyperparticles across dimensions, since the scaling relation
becomes increasingly accurate as $d$ increases, as discussed
in Sec. \ref{approx}.

\begin{table}[!htp]
\caption{Percolation threshold $\eta_c$ of certain overlapping
convex particles $K$ with random orientations in $\mathbb{R}^3$, estimated
using Eq. (\ref{scaling}) and the associated threshold values
$\eta^*_c$ for regular polyhedra (obtained from our numerical simulations) 
and spheroids \cite{ellip2}. 
The threshold of overlapping spheres \cite{paper2, ziff2} is also given, which is
employed to compute the estimates.}
\begin{tabular}{c|c|c}
$K$~~ &  ~~$\eta^*_c$~~ & ~~$\eta_c$~~    \\
\hline
Sphere~ & ~~0.3418~~ & ~~~ \\
Tetrahedron~ & ~~0.1701~~ & ~~0.1774~ \\
Icosahedron~ & ~~0.3030~~ & ~~0.3079~ \\
Decahedron~ & ~~0.2949~~ & ~~0.2998~ \\
Octahedron~ & ~~0.2514~~ & ~~0.2578~ \\
Cube~ & ~~0.2443~~ & ~~0.2485~ \\
Oblate spheroid $a=c=100b$~ & ~~0.01255~~ & ~~0.01154~ \\
Oblate spheroid $a=c=10b$~ & ~~0.1118~~ & ~~0.104~ \\
Oblate spheroid $a=c=2b$~ & ~~0.3050~~ & ~~0.3022~ \\
Prolate spheroid $a=c=b/2$~ & ~~0.3035~~ & ~~0.3022~ \\
Prolate spheroid $a=c=b/10$~ & ~~0.09105~~ & ~~0.104~ \\
Prolate spheroid $a=c=b/100$~ & ~~0.006973~~ & ~~0.01154~ \\
Parallelpiped $a_2=a_3=2a_1$~ & ~~~ & ~~0.2278~ \\
Cylinder $h=2a$~ & ~~~~ & ~~0.4669~ \\
Spherocylinder $h=2a$~ & ~~~~ & ~~0.2972~ \\
\end{tabular}
\label{tab_scaling_3d}
\end{table}

Estimates of the percolation threshold
$\eta_c$ using the scaling relation (\ref{scaling}) for
a variety of randomly oriented overlapping particles in
$\mathbb{R}^3$ are presented in Table \ref{tab_scaling_3d}.
These predictions are compared to
numerically determined threshold values. We see that for most of the randomly oriented
nonspherical particles, the scaling relation (\ref{scaling}),
which is also an upper bound [cf. (\ref{upper})], only
slightly overestimates $\eta_c$ compared with the numerical
results. For  oblate spheroids and prolate spheroid of
aspect ratio 2, the scaling relation slightly underestimates
$\eta_c$ compared to the numerical values, implying that the
numerical results are overestimates of the actual 
threshold for these systems, i.e., the numerical estimates
in these instances violate the upper bound (\ref{upper}).
These three-dimensional results show that scaling relation  (\ref{scaling}) for $\eta_c$ improves
in accuracy over the two-dimensional results reported in Table \ref{tab_scaling_2d}.

\subsubsection{Zero-Volume Convex Plates}

Table \ref{tab_scaling_2din3d} lists the percolation threshold
$\eta_c$ obtained from the scaling relation (\ref{scaling}) 
and numerical simulations for
a variety of randomly oriented overlapping zero-volume two-dimensional plates in
$\mathbb{R}^3$. As per relation (\ref{plate_vol}), the effective volume of 
these plates is chosen to be the
volume of a sphere whose radius $r$ is a characteristic
length scale of the associated plate. In particular, for circular disks, $r$ is chosen to
be radius of the disk. For  noncircular plates,
$r$ is the radius of a disk whose area is the same as that of
the noncircular plate of interest \cite{2din3d}. Importantly, the reference
system for the scaling relation is randomly oriented overlapping
circular disks, which is also given in the table. Again, the scaling relation (\ref{scaling}) only slightly
overestimates $\eta_c$  in the case of the triangular and elliptical plates. For
 square plates, the scaling relation slightly
underestimates $\eta_c$, suggesting that the numerical results overestimate the
actual percolation threshold for this system.

\begin{table}[!htp]
\caption{Percolation threshold $\eta_c$ of certain overlapping
convex plates $K$ with random orientations in $\mathbb{R}^3$, estimated using
Eq. (\ref{scaling}). The associated threshold values $\eta^*_c$
obtained from previous numerical simulations \cite{2din3d} are also given if
available. The threshold of randomly oriented overlapping
circular disks is also given, which is employed to compute the
estimates. The numerical values of $\eta^*_c$ listed here are given 
up to the number of significant figures reported in Ref. \onlinecite{2din3d}. }
\begin{tabular}{c|c|c}
$K$~~ &  ~~$\eta^*_c$~~ & ~~$\eta_c$~~    \\
\hline
Circular disk~ & ~~0.9614 ~~ & ~~~ \\
Square plate~ & ~~0.8647 ~~ & ~~0.8520~ \\
Triangular plate~ & ~~0.7295~~ & ~~0.7475~ \\
Elliptical plate $b=3a$ ~ & ~~ 0.735 ~~ & ~~0.7469~ \\
Rectangular plate $a_2=2a_1$ ~ & ~~ ~~ & ~1.0987~ \\
\end{tabular}
\label{tab_scaling_2din3d}
\end{table}

\subsection{Dimensions Four Through Eleven}

\subsubsection{Nonzero-Volume Convex Bodies}

In Table \ref{tab_eta_hyper}, we provide the percolation threshold
$\eta_c$ of $d$-dimensional overlapping hyperspheres and oriented
hypercubes for $4\le d\le 11$ obtained previously via the rescaled-particle
method \cite{paper2}. As expected, the threshold
values for the two systems approach each other as $d$ increases
and become identical in the limit $d \rightarrow
\infty$, i.e., $\eta_c \rightarrow 1/2^d$ \cite{To12a}. The results for
hyperspheres will be used to estimate $\eta_c$ for the 
hyperparticles discussed in Sec. IV.

\begin{table}[!htp]
\caption{Numerical estimates of the percolation threshold $\eta_c$ of
$d$-dimensional overlapping hyperspheres and oriented hypercubes
for $4\le d\le 11$ reported in Ref. \onlinecite{paper2}. The threshold
values for the two systems,  given up to the number of significant figures reported in Ref. \onlinecite{paper2}, approach each other as $d$ increases
and become identical in the asymptotic limit $d \rightarrow
\infty$, i.e., $\eta_c \rightarrow 1/2^d$ \cite{To12a}. }
\begin{tabular}{c|c|c}
Dimension ~~ &  ~~Hypersphere~~ & ~~ Hypercube ~ \\
\hline
$d$=4 ~ & ~ 0.1304 ~ & ~ 0.1201 ~ \\
$d$=5 ~ & ~ 0.05443 ~ & ~ 0.05024 ~ \\
$d$=6 ~ & ~ 0.02339 ~ & ~ 0.02104 ~ \\
$d$=7 ~ & ~ 0.01051 ~ & ~ 0.01004 ~ \\
$d$=8 ~ & ~ 0.004904 ~ & ~ 0.004498 ~ \\
$d$=9 ~ & ~ 0.002353 ~ & ~ 0.002166 ~ \\
$d$=10 ~ & ~ 0.001138 ~ & ~ 0.001058 ~ \\
$d$=11 ~ & ~ 0.0005530 ~ & ~ 0.0005160 ~ \\
\end{tabular}
\label{tab_eta_hyper}
\end{table}

\begin{table}[!htp]
\caption{Percolation threshold $\eta_c$ of certain $d$-dimensional
randomly overlapping hyperparticles
estimated using Eq. (\ref{scaling}) for $4 \le d \le 11$, including
hypercubes (HC), hyperrectangular
parallelpiped (HRP) of aspect ratio 2 (i.e., $a_1=2a$ and $a_i=a$ for $i=2,\ldots,d$), 
hyperspherocylinder (HSC) of aspect ratio 2 (i.e., $h=2a$), hyperoctahedra (HO)
and hypertetrahedra (HT).}
\begin{tabular}{c|c|c|c|c|c}
Dimension ~~&  ~HC~ & ~HRP~ & ~HSC ~ & ~HO~ & ~HT~  \\
\hline
$d=4$~~ & ~~$8.097\times10^{-2}$~~ & ~~$7.452\times10^{-2}$~~ &
~~$1.109\times10^{-1}$~~ & ~~$6.009\times10^{-2}$~~&
~~$3.471\times10^{-2}$~~ \\
$d=5$~~ & ~~$2.990\times10^{-2}$~~ & ~~$2.775\times10^{-2}$~~ &
~~$4.599\times10^{-2}$~~ & ~~$1.724\times10^{-2}$~~&
~~$8.808\times10^{-3}$~~ \\
$d=6$~~ & ~~$1.167\times10^{-2}$~~ & ~~$1.092\times10^{-2}$~~ &
~~$1.975\times10^{-2}$~~ & ~~$5.560\times10^{-3}$~~&
~~$2.580\times10^{-3}$~~ \\
$d=7$~~ & ~~$4.846\times10^{-3}$~~ & ~~$4.568\times10^{-3}$~~ &
~~$8.899\times10^{-3}$~~ & ~~$1.986\times10^{-3}$~~&
~~$8.518\times10^{-4}$~~ \\
$d=8$~~ & ~~$2.116\times10^{-3}$~~ & ~~$2.006\times10^{-3}$~~ &
~~$4.167\times10^{-3}$~~ & ~~$7.659\times10^{-4}$~~&
~~$3.071\times10^{-4}$~~ \\
$d=9$~~ & ~~$9.584\times10^{-4}$~~ & ~~$9.133\times10^{-4}$~~ &
~~$2.007\times10^{-3}$~~ & ~~$3.129\times10^{-4}$~~&
~~$1.180\times10^{-4}$~~ \\
$d=10$~~ & ~~$4.404\times10^{-4}$~~ & ~~$4.214\times10^{-4}$~~ &
~~$9.746\times10^{-4}$~~ & ~~$1.314\times10^{-4}$~~&
~~$4.696\times10^{-5}$~~ \\
$d=11$~~ & ~~$2.044\times10^{-4}$~~ & ~~$1.963\times10^{-4}$~~ &
~~$4.754\times10^{-4}$~~ & ~~$5.632\times10^{-5}$~~&
~~$1.917\times10^{-5}$~~ \\
\end{tabular}
\label{tab_scaling_hyper}
\end{table}

 To our knowledge,
no numerical estimates of $\eta_c$ are available for 
hyperparticles for $d \ge 4$.  Simulations are
difficult to carry out due to the increasing complexity in
accurately detecting particle overlap as $d$ increases.
Given the accuracy of the scaling relation (\ref{scaling}), verified by numerical results
in $\mathbb{R}^2$ and $\mathbb{R}^3$, we now apply it to estimate
$\eta_c$ for the $d$-dimensional randomly oriented overlapping
hyperparticles discussed in Sec. IV, where it should provide even
better predictions. Table \ref{tab_scaling_hyper} summarizes
threshold values for various hyperparticles for $4 \le d \le 11$
using (\ref{scaling})) and the thresholds for hyperspheres listed in Table \ref{tab_eta_hyper}.
As expected, $\eta_c$ for randomly oriented hypercubes is lower
than that for oriented hypercubes in the corresponding dimensions.

\begin{figure}
\begin{center}
\includegraphics[height=7.5cm,keepaspectratio,clip=]{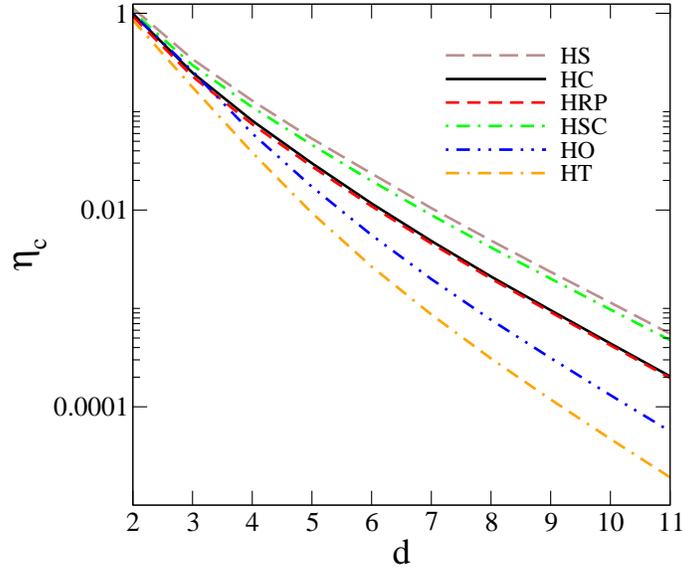}
\end{center}
\caption{(Color online) 
Percolation threshold $\eta_c$ versus dimension $d$ as obtained
from scaling relation (\ref{scaling}) for  randomly oriented
overlapping nonspherical hyperparticles, including hypercubes (HC),
hyperrectangular parallelpipeds (HRP) of aspect ratio 2 (i.e., $a_1=2a$ and $a_i=a$ for $i=2,\ldots,d$), 
hyperspherocylinder (HSC) of aspect ratio 2 (i.e., $h=2a$),
hyperoctahedra (HO) and hypertetrahedra (HT). Values of $\eta_c$ for 
hyperspheres (HS) \cite{paper2} are also shown for purposes of comparison.}
\label{hypers}
\end{figure}

The threshold $\eta_c$ versus $d$ for the systems listed in Table \ref{tab_scaling_hyper} 
is plotted Fig. \ref{hypers}.
As expected, threshold values of nonspherical hyperparticles
are always below that of hyperspheres for any fixed $d$.
For hyperspherocylinders of aspect ratio 2 (i.e., $h=2a$),
the associated $\eta_c$ is only slightly below the hypersphere value
in the  corresponding dimensions. Similarly, for hyperrectangular
parallelpipeds of aspect ratio 2 (i.e., $a_1=2a$ and $a_i=a$ for $i=2,\ldots,d$), 
$\eta_c$ is only slightly below the hypercube value in the
corresponding dimensions.

To obtain excellent estimates of $\eta_c$ for nonspherical
hyperparticles for $d \ge 12$, one can use the scaling relation
(\ref{scaling}) together with highly accurate analytical
expressions for hyperspheres in high dimensions given
in Ref. \onlinecite{To12a}; see, e.g., Eqs. (86)
and (119) in that paper.

\subsubsection{Zero-Volume Convex Hyperplates}

\begin{figure}
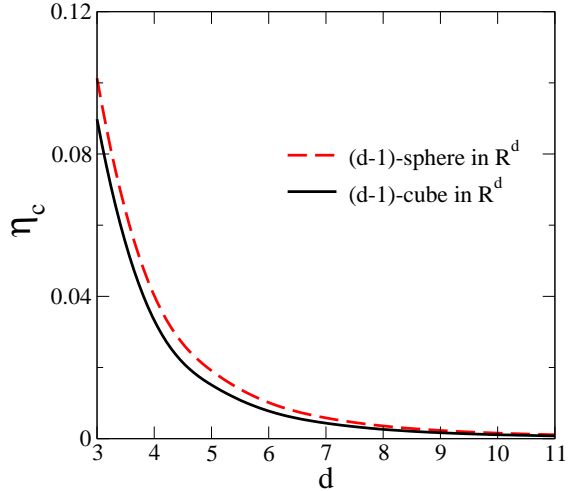

\begin{center}
\includegraphics[height=6.5cm,keepaspectratio,clip=]{Vex_V.eps}
\hspace{.3in}\includegraphics[height=6.5cm,keepaspectratio,clip=]{eta_c_plate.eps}
\end{center}
\caption{(Color online) 
Left panel: Dimensionless exclusion volume $v_{\mbox{\scriptsize ex}}/v_{\mbox{\scriptsize eff} }$ versus dimension $d$ 
for spherical and cubical hyperplates. Right panel: Lower bounds on the percolation threshold $\eta_c$ versus dimension $d$ 
for spherical and cubical hyperplates. }
\label{plates}
\end{figure}

The left panel of Fig. \ref{plates} shows the dimensionless exclusion volume
$v_{\mbox{\scriptsize ex}}/v_{\mbox{\scriptsize eff} }$ versus $d$
for spherical and cubical hyperplates. Recall that the effective
volume $v_{\mbox{\scriptsize eff} }$ of a hyperplate [i.e., a $(d-1)$-dimensional object in
$\mathbb{R}^d$] is chosen to be the volume of a $d$-sphere with radius
$r$ that is a characteristic length  scale of
the hyperplate. For  spherical hyperplates, $r$ is chosen to
be radius of the hyperplate. For  cubical hyperplates,
$r$ is the radius of a $(d-1)$-sphere whose $(d-1)$-dimensional volume
(or $d$-dimensional area) is the same as that of
the cubical hyperplate. Clearly,
$v_{\mbox{\scriptsize ex}}/v_{\mbox{\scriptsize eff} }$ for spherical
hyperplates is smaller than that for cubical hyperplates in all
dimensions $d\ge3$. Lower bounds on $\eta_c$
for spherical and cubical hyperplates, as obtained
from inequality (\ref{aniso-2}), are shown
in the right panel of Fig. \ref{plates}.  As expected, the lower bound for
spherical hyperplates is always greater than that for cubical
hyperplates in all dimensions $d\ge3$.

\section{Conclusions and Discussion}
\label{conc}

We have exploited the principle that low-dimensional continuum percolation 
behavior encodes high-dimensional information, recent analytical results on the percolation threshold $\eta_c$
of continuum percolation models \cite{To12a} and conjecture (\ref{conj}) to formulate the scaling relation (\ref{scaling}) for
$\eta_c$ that is applicable to a wide variety  of overlapping nonspherical 
hyperparticles with random orientations for any dimension $d$. This scaling relation, which is also
an upper bound on $\eta_c$, depends on the $d$-dimensional exclusion volume 
$v_{\mbox{\scriptsize ex}}$ associated with a hyperparticles and was determined 
analytically for, among other shapes, various polygons for $d=2$, Platonic solids, 
spherocylinders, parallepipeds and zero-volume plates for $d=3$ and their appropriate 
generalizations for $d \ge 4$. The scaling relation already provides a good estimate
of $\eta_c$ for $d=2$, becomes increasingly accurate
as $d$ increases, and becomes exact in the high-$d$ limit.
The new estimates of the percolation
thresholds obtained from the scaling relation could have implications 
for a variety of chemical and physical phenomena involving clustering 
and percolation behavior described in the Introduction.  
It will be interesting to see if our conjecture that overlapping hyperspheres provide the highest threshold
among all overlapping systems of convex hyperparticle shapes for any $d$ [cf. (\ref{conj})]
can be proved.

The exclusion volume $v_{\mbox{\scriptsize ex}}$ also has physical importance in other fields besides
percolation, such as granular materials, packing problems, and phase
transitions in condensed-matter systems. For example, $v_{\mbox{\scriptsize ex}}$ 
plays a critical role in the well-known isotropic-nematic
transition in hard-rod systems studied by Onsager \cite{On44}. Thus, our comprehensive 
study of $v_{\mbox{\scriptsize ex}}$ could also deepen our understanding of 
a variety of other problems in condensed matter physics. The relevance
of $v_{\mbox{\scriptsize ex}}$ to packing problems is briefly discussed below.

\begin{figure}
\begin{center}
\includegraphics[height=6.cm,keepaspectratio]{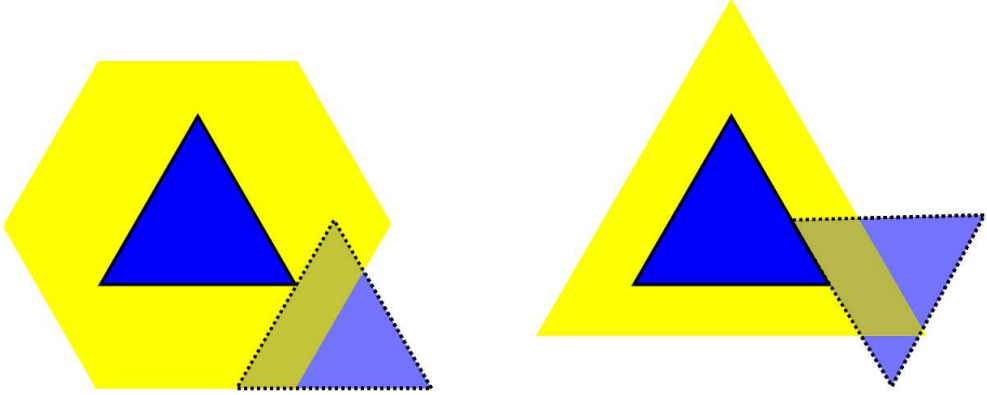}
\end{center}
\caption{(Color online) Exclusion region associated with equilateral triangles 
with volume $v$ with specific relative particle orientations in $\mathbb{R}^2$. Left panel: Triangles are 
perfectly aligned. The associated exclusion region is a regular hexagon (yellow region) whose 
edge length is the same as that of the triangle, resulting in $v_{ex}/v=6$. Right panel: Triangles are 
oriented in such a way that one can be mapped to another by a center inversion operation.
Here two particles always form edge-to-edge contacts. 
The associated exclusion region is a larger equilateral triangle (yellow region) whose 
edge length is twice that of the original triangle, resulting in $v_{ex}/v=4$.}
\label{fig_tri}
\end{figure}

Although the main focus of this paper concerned the percolation
threshold of {\it randomly} oriented 
overlapping hyperparticles, the bound (\ref{aniso}) and the scaling relation (\ref{scaling})
are also valid for overlapping particles with specific orientational 
distributions when the general relation (\ref{v_ave}) for the exclusion volume $v_{ex}$
is used. To illustrate this point, we appeal to two-dimensional examples
depicted in Fig. \ref{fig_tri}, which shows the exclusion regions associated with 
equilateral triangles in $\mathbb{R}^2$ with two different particle orientations: 
fully aligned  and edge-to-edge . Equilateral triangles are not centrally symmetric
and so its exclusion volume $v_{ex}$ is minimized for the edge-to-edge configuration
(i.e., $v_{ex}/v = 4$) and maximized when the particles are aligned 
(i.e., $v_{ex}/v = 6$). The exclusion volume for other relative orientations
must be between these two extremal values. For example, we see from Table \ref{2},
$v_{ex}/v=5.30797\ldots$ for randomly oriented equilateral
triangles. Thus, in order of increasing values of  $v_{ex}/v$ for these
three examples, we have the lower bounds $\eta_c \ge 0.25$ (edge-to-edge) , $\eta_c  \ge 0.188395\ldots$
(random),
and $\eta_c \ge 0.166666\ldots$ (aligned), respectively, as predicted by (\ref{aniso}). The actual threshold
trends for these three cases should follow those given by these bounds,
as does the corresponding predictions of the scaling relation:
$\eta_c \approx 1.1281$, $\eta_c  \approx 0.8501$,
and $\eta_c \approx 0.7520$.

It is noteworthy that the magnitude of the dimensionless exclusion volume 
$v_{\mbox{\scriptsize ex}}/v$ of a convex body with some specified
orientation distribution is closely related to how densely such 
{\it nonoverlapping} objects can be packed in
space \cite{ToJi12}. In particular,  the ratio $v_{\mbox{\scriptsize ex}}/v$ for a centrally symmetric convex hyperparticle 
is minimized when the bodies are aligned \cite{To12a}. For noncentrally
symmetric convex shapes, nonaligned orientations  result in a smaller value
of $v_{\mbox{\scriptsize ex}}/v$ relative to that of the aligned case
(see Fig. \ref{fig_tri}). These exclusion-volume properties of centrally and
noncentrally symmetric bodies play a major role in recently
proposed organizing principles concerning optimal (i.e., maximally
dense) packings of both hard convex and hard concave particles in
$\mathbb{R}^3$ \cite{ToJi12}. Among other results, it was conjectured that
the optimal packing of a centrally symmetric particle is achieved
by the associated optimal Bravais-lattice packing in which all the
particles are aligned (e.g., superballs \cite{superball}); while the optimal packing of a particle
without central symmetry (e.g., truncated tetrahedra \cite{trun_tetrah}) is generally given by a non-Bravais-lattice packing
in which the particles have different nonaligned orientations.

In the Introduction, we discussed the duality between the
equilibrium hard-hyperparticle fluid system and the
continuum percolation model of overlapping hyperparticles
of the same shape.  A consequence of this duality
relation and the so-called decorrelation principle
for disordered hard-hyperparticle packings \cite{To06,Za11} led to the result
that the large-$d$ percolation threshold $\eta_c$ of
overlapping hyperparticles is directly related to the
large-$d$ {\it freezing-point} density (onset of disorder-order phase transition)
of corresponding
equilibrium hard-hyperparticle models \cite{To12a}.
 In a future work, we shall explore this duality
relationship to predict the onset of phase transitions
in hard-hyperparticle fluids across dimensions.
Finally, we will show elsewhere
that the lower-order Pad{\' e] approximants studied in Ref. \onlinecite{To12a} lead also
to bounds on the percolation threshold for lattice-percolation
models in arbitrary dimension.

\section*{Acknowledgements} \vspace{-0.2in} This work was supported by the Materials
Research Science and Engineering Center  Program of the National
Science Foundation under Grant No. DMR-0820341 and by
the Division of Mathematical Sciences at the National Science
Foundation under Award No. DMS-1211087. S. T. gratefully acknowledges the support of a Simons
Fellowship in Theoretical Physics, which has made his sabbatical leave this entire
academic year possible.

\appendix

\renewcommand{\theequation}{A-\arabic{equation}} 
\setcounter{equation}{0}  
\section*{Appendix A: Explicit Calculation of the Mean Width $\bar{w}$ of
A Three-Dimensional Spherical Hyperplate in $\mathbb{R}^4$}
\label{appA}

In Sec. II, we presented an explicit method to calculate the mean
width $\bar{w}=2{\bar R}$ of a convex body in $\mathbb{R}^d$. Recall that the
width $w({\bf n})$ of a convex body is the smallest distance between
two impenetrable parallel $(d-1)-$dimensional hyperplanes contacting
the boundaries of the body, where ${\bf n}$ is a unit vector orthogonal
to the hyperplanes. The mean width $\bar{w}$ is the average of the
width $w({\bf n})$ of the body such that ${\bf n}$ is uniformly distributed
over the unit sphere $S^{d-1}\in\mathbb{R}^d$. Here, we employ this
method to obtain $\bar{w}$ of a three-dimensional spherical hyperplate  (i.e., 3-sphere) of
radius $a$ in $\mathbb{R}^4$.

Due to the rotational symmetry of the 3-sphere, its
orientation in $\mathbb{R}^4$ is specified by a unit vector ${\bf u}$,
which is analogous to the case of a circular disk in $\mathbb{R}^3$; see Table \ref{vol-3-0}.
The width $w({\bf n})$ of the 3-sphere is given by
\begin{equation}
\label{eq_A1}
w({\bf n}) = 2a \sqrt{1-<{\bf u}, {\bf n}>^2} = 2a \sin\theta,
\end{equation}
where $<{\bf x},{\bf y}>$ is the inner product of two vectors ${\bf x}$ and ${\bf y}$, and $\theta$ is the angle
between ${\bf u}$ and ${\bf n}$. Therefore, the mean width is given by
\begin{equation}
\label{eq_A2}
\bar{w} = \frac{1}{\Omega}\int 2a\sin\theta ~dS^3 = \frac{1}{\Omega}\int_0^{2\pi}\int_0^\pi \int_0^\pi 2a\sin^2\theta \sin \phi ~d\theta d\phi d\psi,
\end{equation}
where
\begin{equation}
\label{eq_A3}
\Omega = \int ~dS^3 = \int_0^{2\pi}\int_0^\pi \int_0^\pi \sin\theta \sin \phi ~d\theta d\phi d\psi = 2\pi^{2},
\end{equation}
is the 4-dimensional solid angle. Carrying out the integration in Eq.~(\ref{eq_A2})
yields the mean width $\bar{w}$ of 3-sphere in $\mathbb{R}^4$, i.e.,
\begin{equation}
\label{eq_A4}
\bar{w} = \frac{1}{2\pi^2}\times\frac{16\pi a}{3} = \frac{8a}{3\pi}. 
\end{equation}

\section*{Appendix B: Rescaled-Particle Method for Estimating $\eta_c$ of
Randomly Oriented Overlapping Platonic Solids}
\label{appB}

In Ref. \onlinecite{paper2}, we devised a highly efficient
rescaled-particle method to estimate $\eta_c$ for $d$-dimensional
hyperspheres and oriented hypercubes, utilizing the tightest lower
bound on $\eta_c$, i.e., the pole of the [2, 1] Pad{\' e}
approximant of the mean cluster size $S$, obtained in Ref. \onlinecite{To12a};
see Eq. (119) in that paper. The basic idea behind this method is
to begin with static configurations of overlapping particles at a
reduced density $\eta_0$ that is taken to be the best lower-bound
value and then to increase $\eta$ by rescaling the particle sizes
until a system-spanning cluster forms. This algorithm enables one to accurately estimate
$\eta_c$ in a computationally efficient manner, even in high
dimensions \cite{paper2}. Here, we adapt the method to investigate
percolation properties of randomly oriented overlapping Platonic
solids, including tetrahedra, icosahedra, dodecahedra, octahedra,
and cubes in $\mathbb{R}^3$. Except for overlapping cubes, whose percolation
properties have been well-studied, the percolation thresholds for
the other polyhedra have heretofore not been investigated.

Initially, a Poisson distribution of a large number of points in
the periodic simulation domain is generated. Each point is then taken to be
the centroid of a polyhedron with a random orientation and a
characteristic length $\ell_0$ (e.g., the diameter of the
associated circumsphere). The initial value of $\ell_0$ is chosen
so that the initial reduced density $\eta_0$ of the system
equals the lower-bound value given by (\ref{aniso}). For each
polyhedron $i$, a near-neighbor list (NNL) is obtained that
contains the centroids of the polyhedra $j$ whose distance
$D_{ij}$ to particle $i$ is smaller than $\gamma \ell_0$
($\gamma>1$). The value of $\gamma$ generally depends on the
specific polyhedron of interest. A good choice for $\gamma$ is 
a value  that yields a  NNL list that only
contains polyhedra that overlap at the  threshold. In
our simulations, we used $\gamma \in [1.25, 1.5]$, depending
on the asphericity of the shapes \cite{nature, prepoly}. Then the
sizes of the polyhedra are slowly and uniformly increased by
increasing $\ell_0$, leading to an increase of the reduced
density by an incrementally small amount $\delta \eta$. After each system
rescaling, the polyhedra in the NNL are checked for overlap using
the separation axis theorem \cite{prepoly} and the largest cluster
is identified. This process is repeated until a
system-spanning cluster forms.

Since static particle configurations and a predetermined NNL are
used, the complexity of cluster identification  is significantly reduced. Because initial configurations
with relatively high reduced density values (i.e., the lower bound
values) are used, the amount of rescaling before the system
percolates is much smaller than that from an initial low-density
configuration. Furthermore, the reduced density $\eta$ can
be incremented by a small amount $\delta \eta$  as $\eta_c$ is
approached, leading to a more accurate estimate of the percolation
threshold.

For each of the Platonic solids, we use several distinctly
different system sizes, i.e., $N = 5000, ~10000, ~50000, ~100000$,
and for each system size, we generate 500 independent realizations
of the overlapping polyhedra. The results are extrapolated by
spline fitting the finite-system-size data in a log-log plot to
obtain the infinite-system-size estimate of $\eta_c$.


\begin{thebibliography}{10}

\bibitem{Za83}
R. Zallen, {\it The Physics of Amorphous Solids} (Wiley, New York, 1983).

\bibitem{Br90}
C. J. Brinker and G. W. Scherer, {\it Sol-Gel Science: The Physics and Chemistry
of Sol-Gel Processing} (Academic, New York, 1990).

\bibitem{St92}
D. Stauffer and A. Aharony, {\it Introduction to Percolation Theory} (Taylor \&
Francis, London, 1992).

\bibitem{Sa94}
M. Sahimi, Applications of Percolation Theory (Taylor and Francis, London,
1994).

\bibitem{To02a}
S. Torquato, Random Heterogeneous Materials: Microstructure and
Macroscopic Properties (Springer-Verlag, New York, 2002).

\bibitem{Bross}
V. Myroshnychenko and C. Brosseau, 
J. Appl. Phys. {\bf 103}, 084112 (2008);
S. El Bouazzaoui, A. Droussi, M. E. Achour, and C. Brosseau, 
J. Appl. Phys. {\bf 106}, 104107 (2009).

\bibitem{To12a}
S. Torquato, J. Chem. Phys. {\bf 136},  054106  (2012).

\bibitem{paper2}
S. Torquato and Y. Jiao, J. Chem. Phys. {\bf 137}, 074106 (2012). 

\bibitem{sub}
Note that the volume of a hyperparticle was denoted by $v_1$
in Ref. \onlinecite{To12a}. In the present paper, we drop
the subscript 1, and simply denote this quantity by $v$.

\bibitem{duality}
For equilibrium hard-particle systems, the total
correlation function $h(r;\eta)$ generally takes on both negative
and positive values depending on the values of the radial distance
$r$ and reduced density $\eta$. One can think of the duality
relation (\ref{dual}) as the mapping that is required to convert a
correlation function for a hard-particle system to a non-negative
pair connectedness function (bounded from above by unity) for the
corresponding overlapping particle system.

\bibitem{central}
A hyperparticle is centrally symmetric if its centroid is a point of
inversion symmetry.

\bibitem{exclude}
Roughly speaking, the exclusion volume associated with a particle is
the volume excluded to the centroid of another particle
under the condition that the particles are impenetrable
to one another. Thus, when a particle centroid is inside
another particle's exclusion volume, the particles
necessarily overlap one another. The reader
is referred to Ref. \onlinecite{To12a} for additional
details.


\bibitem{To86i}
S. Torquato, J. Stat. Phys. {\bf 45},  843  (1986).

\bibitem{Za11f}
C.~E. Zachary and S. Torquato, Phys. Rev. E {\bf 84},  056102  (2011).

\bibitem{N_c}
Balberg et al. \cite{Ba84} have
suggested that the mean number of overlaps per particle at the threshold
${\cal N}_c$ is an approximant invariant for overlapping
convex particles of general shape in two and three dimensions.
Simulations have shown this not to be an invariant in
these low dimensions. However, the asymptotic result (\ref{aniso_bd2})
reveals that ${\cal N}_c$ is an invariant, with value unity, in the high-dimensional
limit, regardless of the shape of the convex particle.

\bibitem{Ba84}
I. Balberg, C. H. Anderson, S. Alexander, and N. Wagner, Phys. Rev. B {\bf 30},
3933 (1984).

\bibitem{St95}
D. Stoyan, W.~S. Kendall, and J. Mecke, {\em Stochastic Geometry and Its
  Applications}, 2nd  ed. (Wiley, New York, 1995).

\bibitem{Bo75}
T. Boubl{\'i}k, Mol. Phys. {\bf 29},  421  (1975).

\bibitem{Kih53}
T. Kihara, Rev. Mod. Phys. {\bf 25},  831  (1953).

\bibitem{Sc93}
R. Schneider {\it Convex Bodies: The Brunn-Minkowski Theory}
(Cambridge University Press, Cambridge, 1993).

\bibitem{Co73}
H. S. M. Coxeter, {\it Regular Polytopes}
(Dover, New York, 1973).


\bibitem{Ha79}
H. Hadwiger, Math. Annalen {\bf 239},  271  (1979).

\bibitem{Bet93}
U. Betke and M. Henk, Monatsh. Math. {\bf 115},  27  (1993).

\bibitem{Hen97}
M. Henk, J. Richter-Gebert, and G.~M. Ziegler,  in {\em Handbook of Discrete
  and Computational Geometry} (CRC Press, New York, 1997).





\bibitem{randcube}
D. R. Baker, G. Paul, S. Sreenivasan, and H. E. Stanley,
Phys. Rev. E {\bf 66}, 046136 (2002).


\bibitem{cube2}
S. Mertens and C. Moore, arXiv.1209.4936 (2012).

\bibitem{ellip1}
W. Xia and M. F. Thorpe, Phys. Rev. A {\bf 38}, 2650 (1988).

\bibitem{ellip}
Y. B. Yi and A. M. Sastry, Proc. R. Soc. Lond. A {\bf 460}, 2353 (2007).

\bibitem{ellip2}
E. J. Garboczi, K. A. Snyder, J. F. Douglas, and M. F. Thorpe,
Phys. Rev. E {\bf 52}, 819 (1995).

\bibitem{ziff0}
J. Quintanilla, S. Torquato and R. M. Ziff, J. Phys. A: Math. Gen. {\bf 33}, L399 (2000).

\bibitem{ziff1}
J. Quintanilla and R. M. Ziff, Phys. Rev. E {\bf 76}, 051115 (2007).

\bibitem{ziff2}
C. D. Lorenz and R. M. Ziff, J. Chem. Phys. {\bf 114}, 3659 (2000).

\bibitem{2din3d}
Y. B. Yi and E. Tawerghi, Phys. Rev. E {\bf 79}, 041134 (2009).











\bibitem{On44}
L. Onsager, Phys. Rev. {\bf 65}, 117 (1944).

\bibitem{ToJi12}
S. Torquato and Y. Jiao, Phys. Rev. E {\bf 86}, 011102 (2012).

\bibitem{superball}
Y. Jiao, F. H. Stillinger, and S. Torquato, Phys. Rev. E {\bf 79}, 041309 (2009).

\bibitem{trun_tetrah}
Y. Jiao and S. Torquato, J. Chem. Phys. {\bf 135}, 151101 (2011).

\bibitem{To06}
S. Torquato and F. H. Stillinger, Experimental Math. {\bf 15}, 307 (2006).

\bibitem{Za11}
C. E. Zachary and S. Torquato, J. Stat. Mech.: Theory and Experiment, P10017 (2011).


\bibitem{nature}
S. Torquato and Y. Jiao, Nature {\bf 460}, 876 (2009).

\bibitem{prepoly}
S. Torquato and Y. Jiao, Phys. Rev. E {\bf 80}, 041104 (2009).


\end{thebibliography}

\end{document}